\pdfoutput=1

\documentclass[11pt,twoside,a4paper,cmspaper,final,collab]{cms-tdr}
\def\svnVersion{195904}\def\cmsCernNo{2013-011}\def\cmsCernDate{\today}\def\cmsMessage{Published in Physics Letters B as \href{http://dx.doi.org/10.1016/j.physletb.2013.06.043}{\doi{10.1016/j.physletb.2013.06.043.}}}

\begin{document}\cmsNoteHeader{HIG-12-013}

\hyphenation{had-ron-i-za-tion}
\hyphenation{cal-or-i-me-ter}
\hyphenation{de-vices}

\RCS$Revision: 144489 $
\RCS$HeadURL: svn+ssh://korytov@svn.cern.ch/reps/tdr2/papers/HIG-12-013/trunk/HIG-12-013.tex $
\RCS$Id: HIG-12-013.tex 144489 2012-08-21 04:53:34Z korytov $
\newlength\cmsFigWidth
\ifthenelse{\boolean{cms@external}}{\setlength\cmsFigWidth{0.47\textwidth}}{\setlength\cmsFigWidth{0.48\textwidth}}
\ifthenelse{\boolean{cms@external}}{\providecommand{\cmsLeft}{top}}{\providecommand{\cmsLeft}{left}}
\ifthenelse{\boolean{cms@external}}{\providecommand{\cmsRight}{bottom}}{\providecommand{\cmsRight}{right}}
\providecommand{\re}{\ensuremath{\cmsSymbolFace{e}}}
\hyphenation{ATLAS}

\newcommand{\Nmass}{{120}}
\newcommand{\Nchannels}{{35}} 
\newcommand{\NiuMax}{{161}} 
\newcommand{\NiuMin}{{152}} 

\newcommand{\SMFourExpLNF}{{110}} 
\newcommand{\SMFourExpHNF}{{600}} 

\newcommand{\SMFourObsLNF}{{110}} 
\newcommand{\SMFourObsHNF}{{600}} 

\newcommand{\SMFourObsLNN}{{110}} 
\newcommand{\SMFourObsHNN}{{600}} 

\newcommand{\SMFourObsLNNN}{{110}} 
\newcommand{\SMFourObsHNNN}{{560}} 

\newcommand{\SMFourMinLocalP}{{$1.5 \times 10^{-3}$}}  
\newcommand{\SMFourMaxLocalZ}{{3}}   
\newcommand{\SMFourMaxZmass}{{126}} 

\newcommand{\SMFourNup}{{4}} 
\newcommand{\SMFourGlobalZfull}{{1.6}} 
\newcommand{\SMFourGlobalPfull}{{0.05}} 

\newcommand{\FPObsA}{{110}}
\newcommand{\FPObsB}{{194}}
\newcommand{\FPObsAA}{{110}}
\newcommand{\FPObsBB}{{188}}
\newcommand{\FPObsGapOneAA}{{124.5}}
\newcommand{\FPObsGapOneBB}{{127}}
\newcommand{\FPObsGapTwoAA}{{147.5}}
\newcommand{\FPObsGapTwoBB}{{155}}
\newcommand{\FPMinLocalP}{{0.003}}
\newcommand{\FPMaxLocalZ}{{2.5}}
\newcommand{\FPMaxZmass}{{125.5}}

\newcommand{\PTmiss}{P_{\mathrm{T}}^{\mathrm{miss}} }

\newcommand{\CLs}{\ensuremath{\mathrm{CL_s}\xspace}}

\cmsNoteHeader{HIG-12-008} 

\title{Searches for Higgs bosons in pp collisions at \texorpdfstring{$\sqrt{s}=7$ and 8\TeV}{sqrt(s)=7 and 8 TeV} in the context of four-generation and fermiophobic models}

\date{\today}

\abstract{
Searches are reported for Higgs bosons in the context of either the standard model
extended to include a fourth generation of fermions (SM4) with masses of up to 600\GeV
or fermiophobic models. For the former, results from three decay modes
($\tau\tau$, $\PW\PW$, and $\cPZ\cPZ$) are combined,
whilst for the latter the diphoton decay is exploited.
The analysed proton-proton collision data correspond to integrated luminosities
of up to 5.1\fbinv at 7\TeV and up to 5.3\fbinv at 8\TeV.
The observed results exclude the SM4 Higgs boson
in the mass range 110--600\GeV at 99\% confidence level (CL),
and in the mass range 110--560\GeV at 99.9\% CL.
A fermiophobic Higgs boson is excluded
in the mass range 110--147\GeV at 95\% CL,
and in the range 110--133\GeV at 99\% CL.
The recently observed boson with a mass near 125\GeV
is not consistent with either an SM4 or a fermiophobic Higgs boson.
}

\hypersetup{%
pdfauthor={CMS Collaboration},%
pdftitle={Searches for Higgs bosons in pp collisions at sqrt(s) = 7 and 8 TeV in the context of four-generation and fermiophobic models},%
pdfsubject={CMS},%
pdfkeywords={CMS, physics, Higgs}}

\maketitle 

\newcommand{\gamgam}{\ensuremath{{\gamma\gamma}}\xspace}
\newcommand{\ptgg}{\ensuremath{p_{\mathrm{T}}^{\gamma\gamma}}\xspace}
\newcommand{\mgg}{\ensuremath{m_{\gamma\gamma}}\xspace}
\newcommand{\mjj}{\ensuremath{m_{jj}}\xspace}
\newcommand{\piTgg}{\ensuremath{{\pi_\mathrm{T}^{\gamma\gamma}}}\xspace}
\newcommand{\ygg}{\ensuremath{{y_{\gamma\gamma}}}\xspace}
\newcommand{\cosThetaGG}{\ensuremath{{\cos\theta_{\gamma\gamma}}}\xspace}
\newcommand{\piT}{\ensuremath{{\pi_\mathrm{T}}}\xspace}
\newcommand{\HqT}{{\textsc{HqT}}\xspace}
\newcommand{\mH}{\ensuremath{m_{\mathrm{H}}\xspace}}
\newcommand{\Et}{\ensuremath{E_\mathrm{T}}}
\newcommand{\met}{\ensuremath{\Et^{\mathrm{miss}}}}

\section{Introduction}
\label{sec:intro}

In the standard model (SM)~\cite{Glashow:1961tr,Weinberg:1967tq,sm_salam}, electroweak symmetry breaking is achieved by introducing a complex scalar doublet,
leading to the prediction of the Higgs
boson (\PH)~\cite{Englert:1964et,Higgs:1964ia,Higgs:1964pj,Guralnik:1964eu,Higgs:1966ev,Kibble:1967sv}.
Precision electroweak measurements
indirectly constrain the SM Higgs boson mass $m_{\PH}$ to be less than 158\GeV~\cite{EWKlimits}.
The direct experimental searches exclude at 95\% confidence level (CL) the SM Higgs boson in the mass range up to 600\GeV,
except for the mass window 122--128\GeV
~\cite{LEPlimits, TEVHIGGS_2010, CMSobsJul2012, ATLASobsJul2012},
where a new particle with a mass near 125\GeV was recently observed
in a combination of searches targeting SM Higgs boson decay modes~\cite{ATLASobsJul2012,CMSobsJul2012}.

Various extensions of the standard model have been proposed, such as the inclusion of a fourth generation of fermions (the SM4 model)~\cite{Frampton:1999xi,He:2001tp,Kribs:2007nz,Holdom:2009rf,Erler:2010sk}
or models with multiple Higgs bosons and modified couplings such that one of the Higgs bosons couples only to vector bosons at tree level (the fermiophobic, FP, benchmark model)~\cite{santos99, akeroyd95, akeroyd96, gunion89, akeroyd98, mele_2010}.
  Both types of model have a major impact on Higgs phenomenology.
In the SM4 context for example, constraints from electroweak data become less restrictive,
allowing the mass range 115--750\GeV at 95\% CL, as long as the mass splitting in the fourth generation is $\mathcal{O}$(50)\GeV~\cite{Kribs:2007nz}.
 Likewise Higgs boson production cross sections and decay branching fractions are strongly affected in both scenarios.
Therefore,  the conclusions regarding the existence (or not) of a Higgs boson based on direct searches that assume the SM are not valid in SM4 or FP scenarios without a proper re-interpretation.
Given that the nature of the new boson near 125\GeV has yet to be determined definitively,
it is appropriate to test alternative interpretations beyond the standard model.

To date, the direct searches for the SM4 Higgs
boson have excluded at 95\%~CL the mass range 121--232\GeV~\cite{Tevatron, HWW2010, Aad:2011qi}.
Previous searches using the diphoton decay at the LEP collider~\cite{LepFP},
the Tevatron collider~\cite{Tevatron}, and the Large Hadron Collider (LHC)~\cite{AtlasFP}
exclude a fermiophobic Higgs boson lighter than 121\GeV at 95\% CL.
Using a combination of decay modes, searches at the LHC~\cite{FPpaper}
have ruled out a fermiophobic Higgs boson in the mass range 110--194\GeV at 95\% CL; the range \FPObsAA--\FPObsBB\GeV is excluded at 99\% CL, with the exception of two gaps from \FPObsGapOneAA--\FPObsGapOneBB\GeV and from \FPObsGapTwoAA--\FPObsGapTwoBB\GeV.

In this paper, we re-interpret and combine the SM Higgs boson searches~\cite{CMSobsJul2012, HWW, HZZ2l2nu, HZZ2l2q}, carried out by the Compact Muon Solenoid (CMS) experiment~\cite{bib-detector}
at the LHC, in the SM4 context. The search is performed in the mass range 110--600\GeV. We also report on a search for a fermiophobic Higgs boson in the mass range 110--150\GeV, in the \gamgam decay mode.
The analysed proton-proton collision data correspond to integrated luminosities of up to {5.1}\fbinv at 7\TeV and
up to {5.3}\fbinv at 8\TeV.

\section{ The SM4 and FP models}

The presence of fourth-generation fermions would have
a significant impact on the effective couplings of the Higgs boson to the
SM particles and, thus, directly affect the Higgs boson production
cross sections and decay branching fractions. Since the couplings of the Higgs boson
to fermions are proportional to their masses, the electroweak loop corrections
with fourth-generation fermions have a non-vanishing effect even for arbitrarily heavy
fermions, although perturbative calculations become unreliable for fermion masses larger than $600$\GeV.

In this analysis, we use the SM4 benchmark recommended
by the LHC Higgs cross section group in Ref.~\cite{YR2}:
$m_{\ell_4} = m_{\nu_4} = m_{\cPqd_4} = 600$\GeV and $m_{\cPqu_4} - m_{\cPqd_4} = ( 50 + 10 \cdot \ln(\mH/115) )$\GeV.
Here $m_{\ell_4}$ and $m_{\nu_4}$ are the masses of the 4$^\textit{th}$ generation charged lepton and neutrino,
while $m_{\cPqu_4}$ and $m_{\cPqd_4}$ are the masses of the 4$^{\textit{th}}$ generation ``up'' and ``down'' quarks.
These masses are not excluded by the direct searches for
heavy fermions~\cite{Chatrchyan:2012yea, CMS:2012ab, Aad:2012bt, Aad:2012xc} and
still allow for perturbative calculations.
The SM4 Higgs boson cross sections and decay branching fractions used in this
analysis include electroweak next-to-leading order (NLO) corrections
~\cite{Passarino:2011kv,Denner:2011vt}.
The next-to-NLO order QCD corrections are taken from Ref.~\cite{Anastasiou:2011qw}.
Below we summarise the effect
of the fourth generation fermions, with the specified masses, on the production
and decay of an SM4 Higgs boson compared with the SM Higgs boson of the same mass.

The square of the effective coupling of an SM4 Higgs boson to gluons ($\Pg$) is increased
by a factor $K_{\Pg\Pg}(\mH)$ that ranges between nine and four for a Higgs boson mass that ranges from 110 to 600\GeV.
This enhancement results from the inclusion of $\cPqu_4$ and $\cPqd_4$ quarks in
the quark loop diagrams associated with the $\PH \to \Pg\Pg$ and $\Pg\Pg \to \PH$ processes.
The square of the effective coupling of an SM4 Higgs boson to W and Z vector bosons (henceforth referred to collectively as V bosons)
becomes about three times smaller, $K_{VV}(\mH)\sim 0.3$, as the amplitudes of the NLO and leading order (LO) contributions are of opposite signs in this case.
A coincidental cancellation of the contributions from \PW\ bosons and heavy fermions (top, $\cPqu_4$, $\cPqd_4$, $\ell_4$) to the loop diagrams responsible for the $\PH \to \Pgg\Pgg$ decay suppresses
the square of the effective coupling to photons by $\mathcal{O}$(100).
The squares of the fermionic ($f$) couplings are enhanced by a factor $K_{ff}(\mH) \sim 1.6$.

The enhancement in the effective couplings to gluons and the suppression of couplings to vector bosons causes gluon fusion production to dominate over the vector boson fusion (VBF) and associated (VH) production mechanisms.
Hence, the last two processes can be neglected in searches for SM4 Higgs bosons,
and are ignored in the search presented in this paper.
The contribution from gluon fusion is rescaled by
the SM4/SM $\mH$-dependent factor $K_{\Pg\Pg}(\mH)$ mentioned above.
The $\PH \to \cPqb\cPqb $ search channel that fully relies on associated
production is not included in this combination. For simplicity, $\PH \to \bbbar$ is denoted as $\PH \to \cPqb\cPqb$, $\PH \to \Pgt^+\Pgt^-$ as $\PH \to \Pgt\Pgt$, etc. Following Ref.~\cite{YR2}, the
 uncertainties
on the gluon fusion cross section for the SM4 model are
assumed to be the same as for the SM Higgs boson and are taken from
Ref.~\cite{Dittmaier:2011ti}.
The change in the Higgs boson decay partial widths modifies the decay branching fractions as follows.
The branching fraction $\mathcal{B}(\PH \to \Pgg\Pgg)$ is suppressed by $\mathcal{O}$(100) with respect to the standard model.
The branching fractions $\mathcal{B}(\PH \to \PW\PW)$ and $\mathcal{B}(\PH \to \cPZ\cPZ)$
are suppressed by approximately a factor of five for low Higgs boson masses for which
the $\PW\PW$ and $\cPZ\cPZ$ partial widths are not dominant. They remain almost unchanged in the mid-range around $\mH \sim 200$\GeV,
where vector boson partial widths are the main contributors to the total width $\Gamma_{\text{tot}}$,
and are about 60\% of the SM Higgs boson values above $\mH \sim 350$\GeV
after the $\PH \to \cPqt\cPqt$ decay channel opens up.
The branching fraction $\mathcal{B}(\PH \to \Pgt\Pgt)$ is affected only slightly,
$\mathcal{O}(20\%)$, in the mass range
where this decay mode is used.
The total width
of the SM4 Higgs boson at high masses, where it is  relevant
for the $\PH \to \cPZ\cPZ \to 4\ell$ (where $\ell$ denotes an electron or a muon) search,
is about 30--50\% of the SM Higgs width, depending on the Higgs boson mass.

Since the $\PH \to \Pgg\Pgg$ channel is so strongly suppressed,
it has nearly no sensitivity for the SM4 Higgs boson and is therefore not included in the combination.
We explicitly checked that including or omitting this
channel has no effect on the combined SM4 Higgs boson search results
even in the presence of the significant excess near 125\GeV observed in the standalone search for $\PH \to \Pgg\Pgg$~\cite{CMSobsJul2012}.

The theoretical uncertainties on the SM4 Higgs boson decay branching fractions are derived
from three independent sources of relative uncertainty on the partial widths,
which amount to approximately 50\%, 10\%, and 5\% for $\Gamma_{VV}$, $\Gamma_{ff}$, and $\Gamma_{\Pg\Pg}$, respectively~\cite{YR2}.
Any given decay channel $\PH \to xx$ is affected by each of these three uncertainties.
Using the equation $\mathcal{B}_{xx} = \Gamma_{xx} / \Gamma_{\text{tot}}$ and standard error propagation,
we translate the uncertainties on the partial widths into uncertainties on the branching fractions
of the decay modes ($\Pgt\Pgt$, $\PW\PW$, $\cPZ\cPZ$) used in this combination.
The signal acceptance for each exclusive final state is assumed to be the same as reported in previous SM Higgs boson
searches~\cite{CMSobsJul2012, HWW, HZZ2l2nu, HZZ2l2q}.

As a fermiophobic Higgs boson does not couple to fermions, gluon fusion production becomes negligible,
while the VBF and VH production cross sections remain unchanged. Direct decays to fermion pairs become impossible,
which significantly increases the branching fractions $\mathcal{B}(\PH \to \Pgg\Pgg)$, $\mathcal{B}(\PH \to \PW\PW)$
and $\mathcal{B}(\PH \to \cPZ\cPZ)$. The diphoton decays are enhanced further as the negative interference between
the W and top loops responsible for this decay in the SM is no longer present.
For a low mass FP Higgs boson ($\mH \approx 125\GeV$) the decay to two photons
is enhanced by an order of magnitude with respect to the SM~\cite{gunion89, akeroyd98, mele_2010},
and this compensates for the reduced production cross section, keeping the overall diphoton signal rate very similar to that in the SM.
Production cross sections and decay branching fractions, together with their uncertainties,
are taken from Ref.~\cite{Dittmaier:2011ti} and are derived from
Refs.~\cite{Arnold:2008rz,Bolzoni:2010xr,Brein:2003wg,Ciccolini:2003jy,Djouadi:1997yw,hdecay2}.

\section{The CMS detector and event reconstruction}

The CMS apparatus~\cite{bib-detector} consists of a barrel assembly and two endcaps,
comprising, in successive layers outwards from the collision region,
the silicon pixel and strip tracker,
the lead tungstate crystal electromagnetic calorimeter (ECAL),
the brass/scintillator hadron calorimeter,
the superconducting solenoid,
and gas-ionization chambers embedded in the steel flux return yoke for the detection of muons.
The polar coordinate system ($\theta$, $\phi$) is used to describe the
direction of particles and jets emerging from the pp collisions, where
$\theta$ is the polar angle measured from the positive $z$ axis (along the anticlockwise beam direction) and $\phi$ is the azimuthal angle.
The pseudorapidity, defined as $\eta=-\ln \left[ \tan(\theta/2) \right]$, is commonly used in place of $\theta$.

Particles are reconstructed with the CMS ``particle-flow'' event description~\cite{CMS-PAS-PFT-09-001,CMS-PAS-PFT-10-001}
using an optimized combination of all subdetector information to form ``particle-flow objects'':
electrons, muons, photons, charged and neutral hadrons.
Jets are formed by clustering these objects with the anti-\kt algorithm~\cite{Cacciari:2008gp}
using a distance parameter $\DR = 0.5$, where
$\DR = \sqrt{(\Delta\eta)^2+(\Delta\phi)^2}$ and
$\Delta\eta$ and $\Delta\phi$ are the pseudorapidity and azimuthal angle
differences between the jet axis and the particle direction.
The missing transverse energy vector, $\vec{E}_{\mathrm{T}}^{\text{miss}} $,
is taken as the negative vector sum of all particle transverse
momenta, and its magnitude is referred to as $\ETmiss$.

\section{Search channels}
\label{sec:analyses}

\subsection{The SM4 search channels}
\label{sec:SM4analyses}

The SM4 results presented are obtained by combining
searches in the individual Higgs boson decay channels listed in Table~\ref{tab:channels}.
The table summarizes the main characteristics of these searches, namely:
the mass range of the search,
the integrated luminosity used,
the number of exclusive sub-channels,
and the approximate instrumental mass resolution.

\begin{table*}[htbp]
\begin{center}
\small
  \topcaption[ ] {Summary of the analyses included in the SM4 combination.}
  \label{tab:channels}
\begin{tabular}{ l  r@{}l c c c r l }
\hline 
\multirow{2}{*}{Channel}                                                & & $\mH$ range   & \multicolumn{2}{c}{Int. lumi. (\!\fbinv)}     & Sub-      & $\mH$       & Ref.            \\
                                                                        & & (\GeVns{})      &    7\TeV   &   8\TeV                  & channels  & resolution  &                 \\
\hline 
$\PH \to \Pgt\Pgt \to \Pe\tau_{\mathrm{h}}/\mu\tau_{\mathrm{h}}/\Pe\mu/\mu\mu$   & & 110--145      & 4.9   &  5.1                     &  16       & 20\%        & \cite{CMSobsJul2012}   \\
$\PH \to \PW\PW \to 2\ell 2\nu$                                              & & 110--600      & 4.9   &  5.1                     &  4        & 20\%        & \cite{HWW, CMSobsJul2012}       \\
$\PH \to \cPZ\cPZ \to 4\ell$                                                 & & 110--600      & 5.0   &  5.3                     &  3        & 1--2\%      & \cite{CMSobsJul2012}     \\
$\PH\to\cPZ\cPZ \to 2\ell 2\nu$                                              & & 250--600      & 4.9   &  --                       &  2        & 7\%         & \cite{HZZ2l2nu}  \\
\multirow{2}{*}{$\PH \to \cPZ\cPZ \to 2\ell 2\cPq$}&\multirow{2}{*}{$\bigg\{$}& 130--164 & \multirow{2}{*}{4.9}&\multirow{2}{*}{--}&\multirow{2}{*}{6} & {3\%}&\multirow{2}{*}{\cite{HZZ2l2q} }\\
 &   &200--600 &    &                         &     &{3\%} &  \\
\hline 
\end{tabular}
\end{center}
\end{table*}

Below we give a brief summary of the individual searches. More detailed
descriptions of all analyses can be found in Refs.~\cite{CMSobsJul2012, HWW, HZZ2l2nu, HZZ2l2q}.
In the combination presented here, Higgs boson production via VBF is neglected,
 and thus sub-channels
in the $\PH \to \Pgt\Pgt$ and $\PH \to \PW\PW$ decay channels that
explicitly target VBF production are also dropped.

The $\PH \to \Pgt\Pgt$ search~\cite{CMSobsJul2012}
is performed using the final-state signatures $\Pe\Pgm$, $\Pgm\Pgm$, $\Pe \Pgt_{\mathrm{h}}$, and $\Pgm \Pgt_{\mathrm{h}}$, where
electrons and muons arise from leptonic $\Pgt$ decays and
$\Pgt_{\mathrm{h}}$ denotes hadronic $\Pgt$ decays.
Each of these categories is further divided into 4 exclusive sub-categories based on the jet multiplicity and
transverse momentum ($\pt$) of the visible tau lepton decay.
In each category, we search for a broad excess in the reconstructed $\Pgt\Pgt$ mass distribution.
The main irreducible background, $\cPZ \to \Pgt\Pgt$ production,
and the largest reducible backgrounds ($\PW+\text{jets}$, multijet production, $\cPZ\to \Pe\Pe$)
are evaluated from various control samples in data.

The $\PH \to \PW\PW \to 2\ell 2\cPgn$ analysis~\cite{HWW, CMSobsJul2012}
searches for an excess of events with two leptons of opposite charge,
large missing transverse energy
$\ETmiss$, and less than two jets.
Events are divided into four categories, with different background compositions
and signal-to-background ratios, according to the number of jets and whether the leptons are of the same or different flavour.
For events with no jets, the main background stems from non-resonant $\PW\PW$ production;
for events with one jet, the dominant backgrounds are from $\PW\PW$ and top-quark production.
The events are split
into same-flavour and different-flavour dilepton sub-channels, since the background from Drell--Yan production is much larger
for the same-flavour dilepton events.
To improve the separation of signal from background in the 7\TeV analysis, multivariate analysis classifiers are
trained for a number of Higgs boson masses,
and a search is made for an excess of events in the output distributions of the classifiers.
All background rates, except for small expected contributions from $\PW\cPZ$, $\cPZ\cPZ$, and $\PW\Pgg$,
are evaluated from data.

In the $\PH \to \cPZ\cPZ \to 4\ell$ channel~\cite{CMSobsJul2012},
we search for a four-lepton mass peak over a small continuum background.
To separate signal and background, we use a discriminant
calculated for each event as the ratio of the respective probability densities for signal and
background to form an event with the observed kinematic
configuration of four leptons.
The $4\Pe$, $4\Pgm$, and $2\Pe2\Pgm$ sub-channels are analysed separately
since there are differences in the four-lepton mass resolutions
and the background rates arising from jets misidentified as leptons.
The dominant irreducible background in this channel is from non-resonant $\cPZ\cPZ$ production
with both $\cPZ$ bosons decaying to either $2\Pe$, $2\Pgm$, or $2\Pgt$ (with the tau leptons decaying leptonically) and is estimated from simulation.
The smaller reducible backgrounds with jets misidentified as leptons, \eg $\cPZ+\text{jets}$, are estimated from data.

In the $\PH\to\cPZ\cPZ  \to 2\ell 2\cPgn$ search~\cite{HZZ2l2nu},
we select events with a lepton pair ($\Pep\Pem$ or $\Pgmp\Pgmm$),
with invariant mass consistent with that of an on-shell $\cPZ$ boson,
and a large missing transverse energy.
We then define a transverse invariant mass $m_{\mathrm{T}}$ from the dilepton momenta
and $\vec E_{\mathrm{T}}^{\mathrm{miss}} $,
 which is assumed to originate from neutrinos in the $\cPZ \to \nu\nu$ decays,
and search for a broad excess of events in the $m_{\mathrm{T}}$ distribution.
The $\cPZ\cPZ$ and $\PW\cPZ$ backgrounds are taken from simulation,
while all other backgrounds, $Z+\text{jets}$ and a cumulative sum of the rest,
are evaluated from control samples in data.

In the $\PH \to \cPZ\cPZ  \to 2\ell 2\cPq$ search~\cite{HZZ2l2q},
we select events with two oppositely-charged leptons ($\Pe^+\Pe^-$ or $\Pgm^+\Pgm^-$),
and two jets. The two leptons and the two jets are required to have invariant masses
consistent with that of on-shell $\cPZ$ bosons. The events are categorized by the lepton flavour and the number of jets identified as coming from the decay of a $\cPqb$-quark, thus defining six exclusive final states. We search for a peak in the invariant mass distribution
of the dilepton-dijet system, with the background rate and shape estimated using control regions in data.

\subsection{The FP search channels}
\label{sec:FPanalyses}

In this section, we describe the FP Higgs boson search with the 8\TeV dataset.
We use the $\PH \to \gamgam$ decay mode and exploit the characteristic signatures
associated with the VBF and VH processes: namely, the two forward jets produced
by the scattered quarks in VBF production and charged leptons (electrons or muons)
or large missing transverse energy induced by neutrinos,
both coming from vector boson decays in VH production.
The FP Higgs boson search in the diphoton decay mode with the 7\TeV dataset
is described elsewhere~\cite{FPpaper}.

The simulated VBF signal samples are generated with {\POWHEG}~\cite{powheg}.
The difference in the event selection acceptance for samples generated with \POWHEG
at NLO and with {\PYTHIA}~\cite{Sjostrand:2006za} at LO is taken as a systematic uncertainty,
which is found to have a negligible impact on the final results.
The simulated VH samples are generated with {\PYTHIA}.

Nine exclusive classes are defined. All require two, isolated, high $\pt$ photons.
Five of the nine require an additional tag: either a pair of jets
(subdivided into two sub-classes with low and high dijet invariant masses, $\mjj$),
or an isolated lepton (subdivided into $\Pe$ and $\mu$ sub-classes),
or a large missing transverse energy. The remaining diphoton events
failing
to pass VBF and VH production tags
form an untagged category, which is divided into four sub-classes according to
the photon shower shape and position in the detector~\cite{CMSobsJul2012}.
The selection criteria for the photon candidates are the same as in the SM search~\cite{CMSobsJul2012} except for the modifications noted below.
 A Higgs boson produced via the VBF or VH mechanisms typically has a larger $\PT$ than a Higgs boson produced via gluon fusion (which
 dominates SM Higgs production) and hence the photon $\pt$ thresholds are increased.
  Furthermore, such photons also have a harder transverse
momentum spectrum than those of photons produced by background
processes~\cite{Ballestrero:2008gf} and thus significant separation of signal and background can be achieved. The transverse
momentum of the photon pair ($\ptgg$) together with their invariant mass ($\mgg$) are included in a
two-dimensional unbinned maximum likelihood. The signal and background models,
 which are used to extract limits on the signal cross section, are described in detail in Ref.~\cite{FPpaper}.
The dijet-tagged class has the greatest sensitivity; here the background model is derived from data, by fitting the diphoton
mass distributions over the range $100 < \mgg < 180$\GeV.

In the dijet-tagged classes the photon $\pt$ thresholds are raised (compared with the SM search~\cite{CMSobsJul2012}) to
$\pt^{\gamma}(1) > \mgg/2$, and $\pt^{\gamma}(2) > 25$\GeV, where $\pt^{\gamma}(1)$ and $\pt^{\gamma}(2)$ are the transverse momenta of the leading and
sub-leading photons respectively.
The $\pt$ thresholds for the two jets are 30\GeV and 20\GeV, and
their separation in $\eta $ must be greater than 3.0.
 The dijet mass is required to be greater than 250\GeV.
 The selected events are subdivided into two regions 250 $ < \mjj <500\GeV$ and $\mjj >  500\GeV$, based on the amount of background contamination as a function of dijet mass.
In addition, for events with $\mjj > $ 500\GeV, the $\pt$ threshold for the subleading jet is raised to 30\GeV. Two additional selection criteria, relating the dijet
and diphoton systems, are applied to all selected events. The difference between the average
$\eta $ of the two jets and the $\eta $ of the diphoton system is required to be
less than 2.5~\cite{Rainwater:1996ud}. The difference in
 $\phi$ between the diphoton and
dijet systems is required to be greater than 2.6 radians.

In the lepton-tagged channel, which targets VH production, the $\pt$ thresholds are again altered; values of
 $\pt^{\gamma}(1)>3\times\mgg/8$, and $\pt^{\gamma}(2)>25\GeV$
 are set. Separate muon and electron sub-classes are defined, with at least one muon (electron) with $\pt> 20\GeV$ and within $|\eta|< 2.4$ ($|\eta|< 2.5$) required.  The leptons must be isolated, using isolation criteria similar to those used for photons, and separated from the photons by
$\Delta R > 1$.
 To protect against background events that arise from an electron misidentified as a photon in the $\cPZ \to \Pe\Pe$ process, the mass of the photon-electron system must differ from the Z boson mass by at least 5\GeV.

A significant fraction of events from VH production contains large missing transverse energy
due to the neutrinos from $\cPZ \rightarrow \nu\nu$ decays.
 Events that passed the requirements of the lepton-tag channel
  are excluded to form a statistically
independent $\met$-tag class. The $\met$ is required to be larger than 70\GeV.
The photon $\pt$ threshold requirements are the same as for the lepton-tag class.
Due to the negligible contribution of photons at large pseudorapidity to the expected exclusion limit,
only photons falling within the ECAL barrel are kept ($|\eta| < 1.48$).

A substantial fraction of the FP signal events are not
expected to pass any of the previous tags, and so
the remaining untagged events are also exploited.
Photon $\pt$ requirements of  $\pt^{\gamma}(1) > \mgg/3$, $\pt^{\gamma}(2) > \mgg/4$ and $\ptgg/\mgg>0.1$ are applied.
 The selected events are divided into four classes
according to the expected mass resolution and amount of background contamination~\cite{CMSobsJul2012}.
Two classifiers are used: the minimum $\RNINE$ of the two photons,
$\RNINE^\text{min}$, and the maximum absolute pseudorapidity of the two photons.
The quantity $\RNINE$ is defined as the sum of the energy in the 3x3 crystal array centred on
the crystal with the maximum energy deposit divided by the
total clustered energy,
 and is designed to identify photons undergoing a conversion.
The untagged diphoton event classes are: (a) both photons in the barrel
and $\RNINE^\text{min} > 0.94$, (b) both photons in the barrel and
$\RNINE^\text{min} < 0.94$, (c) one or both photons in the endcaps and
$\RNINE^\text{min} > 0.94$, and (d) one or both photons in the endcaps and
$\RNINE^\text{min} < 0.94$.

\begin{table*}[htbp]
\begin{center}
\topcaption{Number of selected events in the \gamgam event classes,
for data in the mass range 100--180\GeV and for a fermiophobic Higgs boson signal ($\mH$ = 125\GeV).
The expected number of background events in the signal region 120--130\GeV obtained from
the fit of the data in the full mass range 100--180\GeV
and the mass resolution for the 125\GeV FP Higgs boson signal in each event class are also given.
All numbers are for the 8\TeV dataset.
}
\begin{tabular}{lcccccccc}
\hline
\multirow{3}{*}{} & $\met$ & Dijet & Dijet & Lepton & \multicolumn{4}{c}{Untagged} \\
\cline{6-9}
& tag & high \mjj & low \mjj & tag ($\Pe,\mu$) & (a) & (b) & (c) & (d) \\
\hline
Data 		          & 41  & 84   & 271  & 30 & 4992 & 9546 & 5105 & 8574 \\
Signal ($\mH = 125\GeV$)             & 2.3 & 14 & 10 & 3.5 & 18 & 23 & 12 & 14 \\
Expected background & 5.8   & 17 & 40 & 4.1  & 740 & 1400 & 760 & 1300 \\
\hline
$\sigma_\text{eff}$ (\GeVns{})  & 2.0  & 2.1  & 2.2 & 2.1  & 1.5 & 2.0 & 3.8 & 3.9 \\
\hline
\end{tabular}
\label{tab:ClassFracs}
\end{center}
\end{table*}

The numbers of events in the
\gamgam event classes
are shown in Table~\ref{tab:ClassFracs}, for simulated signal events and for data.
A Higgs boson with $\mH$ = 125\GeV is chosen for the signal, and the
data are counted in the mass range 100--180\GeV.
The table also shows the mass resolution, $\sigma_\text{eff}$,
defined as half the width of the narrowest window containing 68.3\% of the
distribution.

\section{Combination method}
\label{sec:method}

The combination of the Higgs boson searches, be it across different sub-channels within a given decay mode
or across different decay modes, requires simultaneous analysis of the data selected by all individual analyses,
accounting for all statistical and systematic uncertainties and their correlations.
The overall statistical methodology used in this combination was developed
by the ATLAS and CMS Collaborations in the context of the LHC Higgs Combination Group.
The description of the general methodology can be found in Refs.~\cite{LHC-HCG-Report, CMScombFeb2012}.
Below we give concise definitions of statistical quantities
we use for characterizing the outcome of the search.
Results presented in this paper are obtained using asymptotic formulae~\cite{Cowan:2010st},
including a few updates recently introduced in the {\sc RooStats} package~\cite{RooStats1}.

For calculations of exclusion limits, we adopt the modified frequentist
criterion $\CLs$~\cite{Junk:1999kv,Read1}. The chosen test statistic, $q_{\mu}$, used to determine
how signal- or background-like the data are, is based on the profile likelihood ratio.
Systematic uncertainties are incorporated in the analysis via nuisance parameters and
are treated according to the frequentist paradigm.
The profile likelihood ratio is defined as

\begin{equation}
q_{\mu} \, = \, - 2 \, \ln \frac {\mathcal{L}(\text{obs} \, | \, \mu \cdot s + b, \, \hat \theta_{\mu} ) }
                           {\mathcal{L}(\text{obs} \, | \, \hat \mu \cdot s + b, \, \hat \theta ) } ,
\end{equation}

where $s$ stands for the expected number of signal events under the SM4/FP Higgs boson hypothesis,
$\mu$ is a signal strength modifier
introduced to accommodate deviations from SM4/FP Higgs boson predictions,
$b$ stands for backgrounds, and $\theta$ are nuisance parameters
describing systematic uncertainties
The likelihood in the numerator reaches its maximum, for a given $\mu$, at $\hat \theta_{\mu}$; while $\hat \mu$ and $\hat \theta$ define the point at which the likelihood reaches its global maximum. The quantity $\hat \mu$ is constrained to be between 0 and $\mu$.

The ratio of probabilities to observe a value of the test statistic
at least as large as the one observed in data, $q_{\mu}^{\text{obs}}$,
under the signal+background (s+b) and background-only (b) hypotheses,

\begin{equation}
\CLs = \frac {\mathrm{P}(q_{\mu} \geq q_{\mu}^{\text{obs}} \, | \, \mu \cdot s+b)}
             {\mathrm{P}(q_{\mu} \geq q_{\mu}^{\text{obs}} \, | \, b )} \, \leq \alpha ,
\end{equation}

is used as the criterion for excluding the signal at the $1 - \alpha$ confidence level.

To quantify the presence of an excess of events over what is expected for the background,
we use another test statistic where the likelihood appearing in the numerator
is for the background-only hypothesis:
\begin{equation}
q_{0} \, = \, - 2 \, \ln \frac {\mathcal{L}(\text{obs} \, | \, b, \, \hat \theta_{0} ) }
                       {\mathcal{L}(\text{obs} \, | \, \hat \mu \cdot s + b, \, \hat \theta ) } ,
\end{equation}

The statistical significance $Z$ of a signal-like excess is computed from the probability $p_0$

\begin{equation}
p_0 = \mathrm{P}(q_0 \geq q_0^{obs} \, | \, \text{b}),
\end{equation}

henceforth referred to as the $p$-value, using the one-sided Gaussian tail convention:

\begin{equation}
\label{eq:Z}
p_0 \, = \, \int_{Z}^{+\infty} \frac{1}{\sqrt{2\pi}} \exp(-x^2/2) \,\, \rd{}x.
\end{equation}

In the Higgs boson search, we scan over Higgs boson mass hypotheses and
look for the one giving the minimum local $p$-value $p_{\text{local}}^{\text{min}}$,
which describes the probability of a background fluctuation for that particular Higgs boson mass
hypothesis. The probability to find a fluctuation with a local $p$-value lower or equal to the observed
$p_{\text{local}}^{\text{min}}$ anywhere in the explored mass range
is referred to as the global $p$-value, $p_{\text{global}}$.

The fact that $p_{\text{global}}$ can be significantly larger than $p_{\text{local}}^{\text{min}}$
is often referred to as the look-elsewhere effect.
The global significance (and global $p$-value) of the observed excess can be evaluated in this case by
generating pseudo-datasets, which, however, becomes too computationally intensive and not practical
for very small $p$-values.
Therefore, we use the method suggested in Ref.~\cite{LEE}. The relationship between
global and local $p$-values is given by:

\begin{equation}
\label{eq:LEE1}
p_{\text{global}} = p_{\text{local}}^{\text{min}} + C \cdot \re^{ - Z^2_{\text{local}} / 2 }
\end{equation}

When the look-elsewhere effect is very large, as in this search,
the constant $C$ can be evaluated directly from data~\cite{LHC-HCG-Report}
by counting upcrossings $N_{\mathrm{up}}$ of $\hat \mu (\mH)$ with the line $\mu=0$ and setting
$C = N_{\mathrm{up}}$. The best-fit signal strength $\hat \mu$ in this case is obtained
from maximizing the likelihood $\mathcal{L}(\text{obs} \, | \, \hat \mu \cdot s + b, \, \hat \theta )$
with no constraints on $\hat \mu$.

\section{Results}
\label{sec:results}

The following conventions are used. The observed values
are shown in the plots by a solid line.
A dashed line is used to indicate the median of the expected results for
the background-only hypothesis.
The green (dark) and yellow (light) bands show the ranges in which
the measured values are expected to reside in at least 68\% and 95\%
of all experiments under the background-only hypothesis.

\subsection{The SM4 results}
\label{sec:sm4results}

\begin{figure}[thbp]
\centering
\includegraphics[width=\cmsFigWidth]{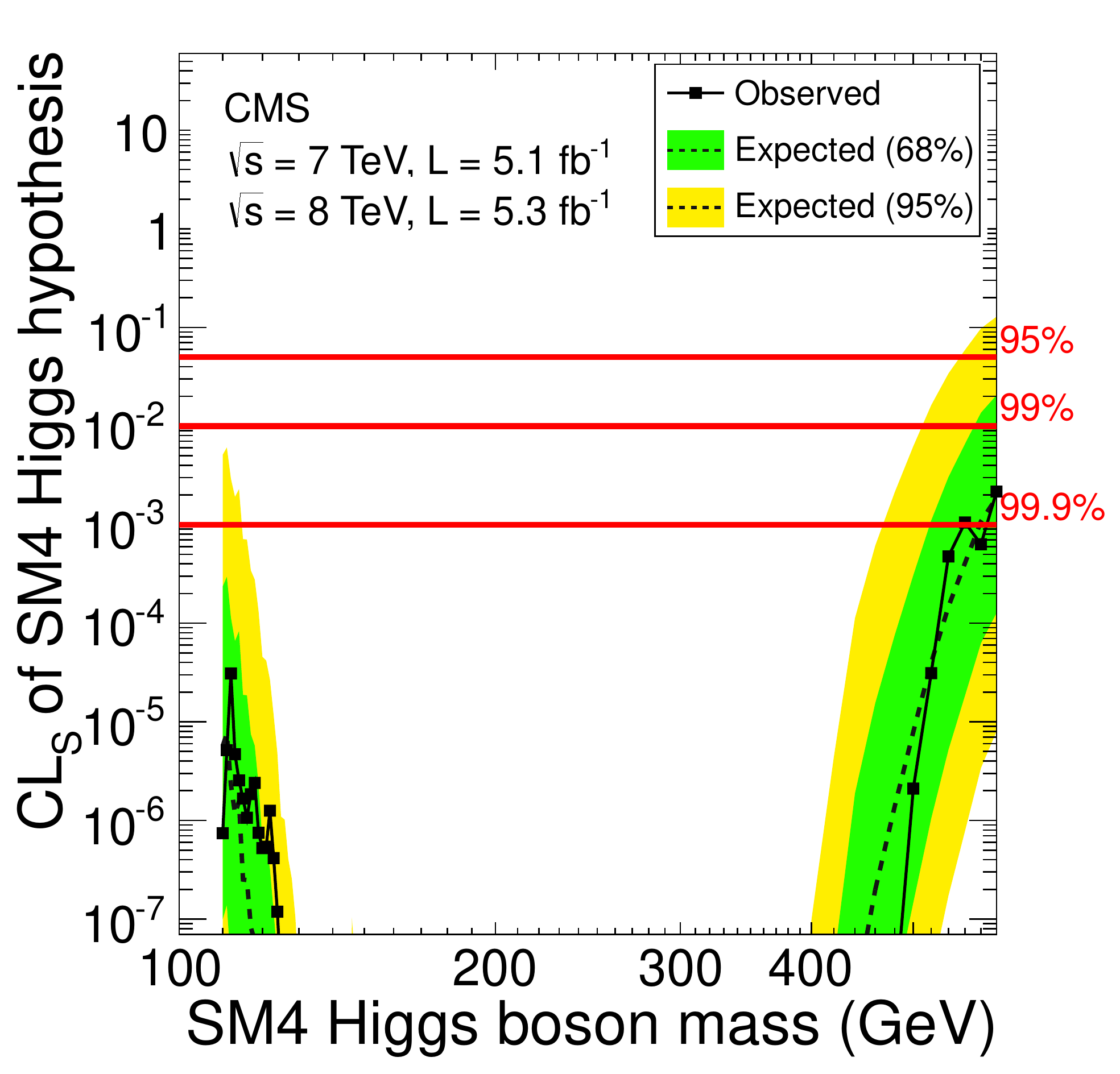}
\includegraphics[width=\cmsFigWidth]{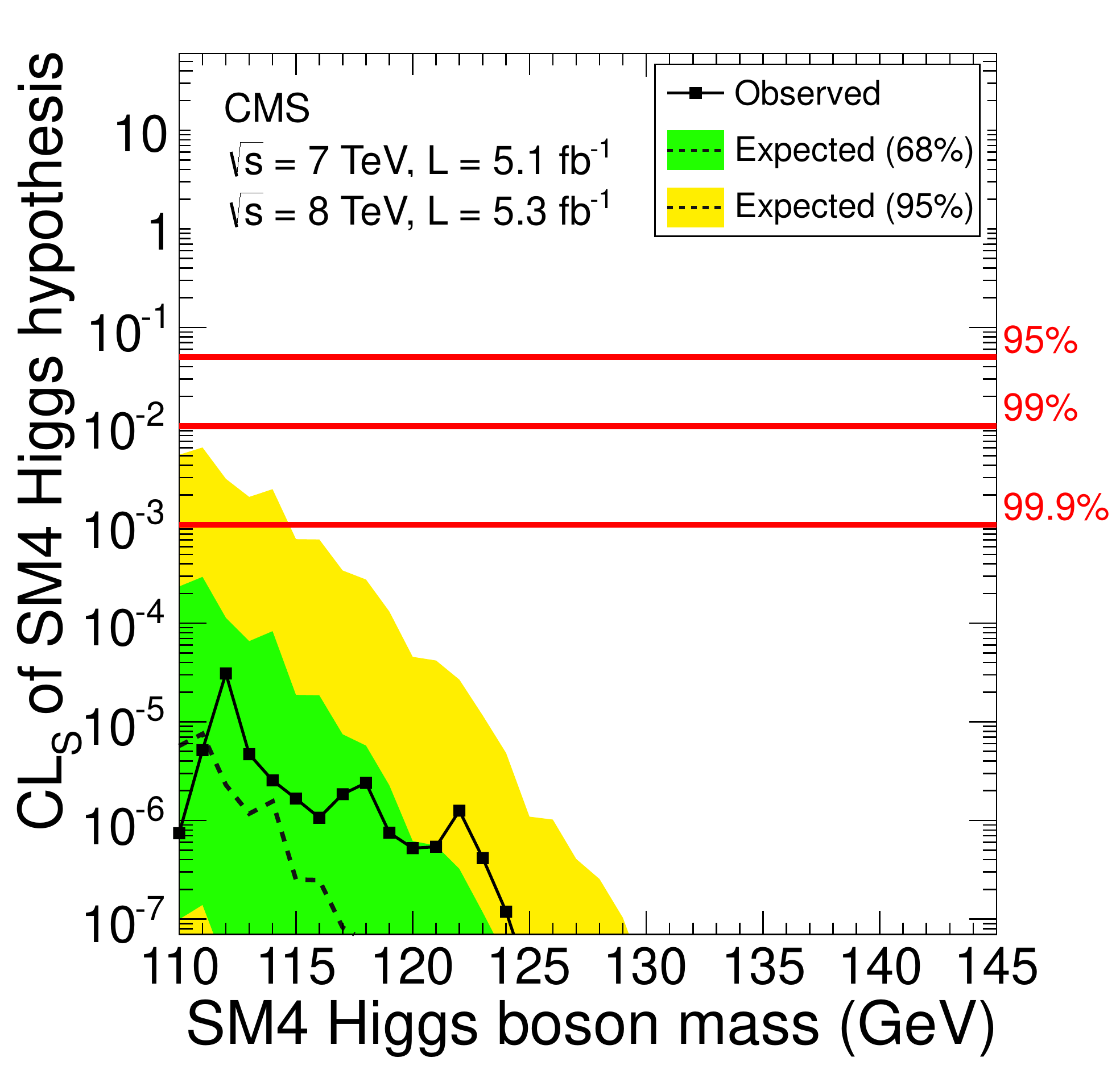}
\caption{
  The { observed and expected} $\CLs$ values for the SM4 Higgs boson hypothesis as a function
  of the Higgs boson mass in the range 110--600\GeV (\cmsLeft) and 110--145\GeV (\cmsRight).
  The three
  horizontal lines show confidence levels of 90\%, 95\%, and 99\%, defined as ${(1-\CLs)}$.
}
\label{fig:CLs}
\end{figure}

\begin{figure} [hbtp]
\centering
\includegraphics[width=\cmsFigWidth]{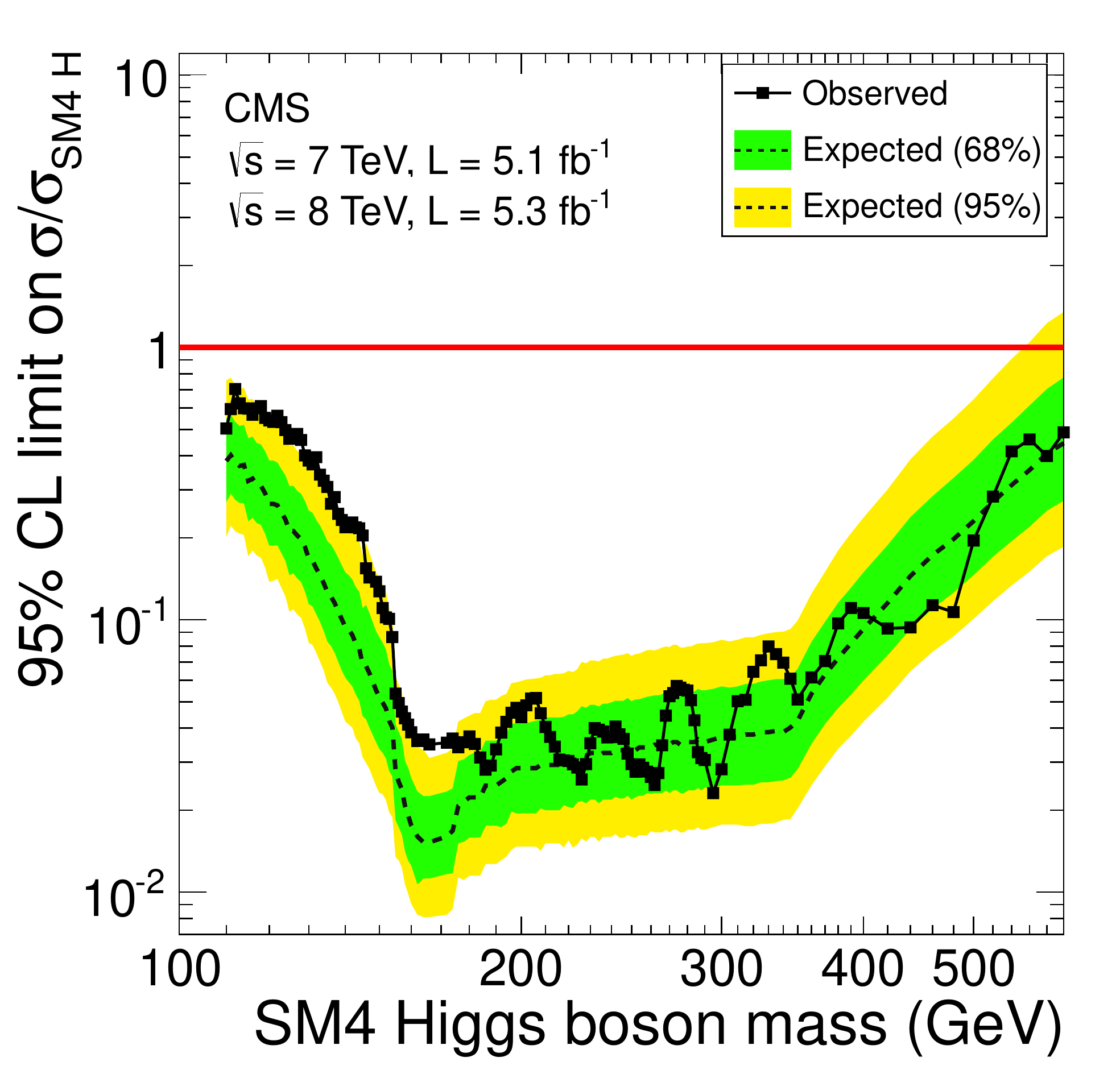}
\includegraphics[width=\cmsFigWidth]{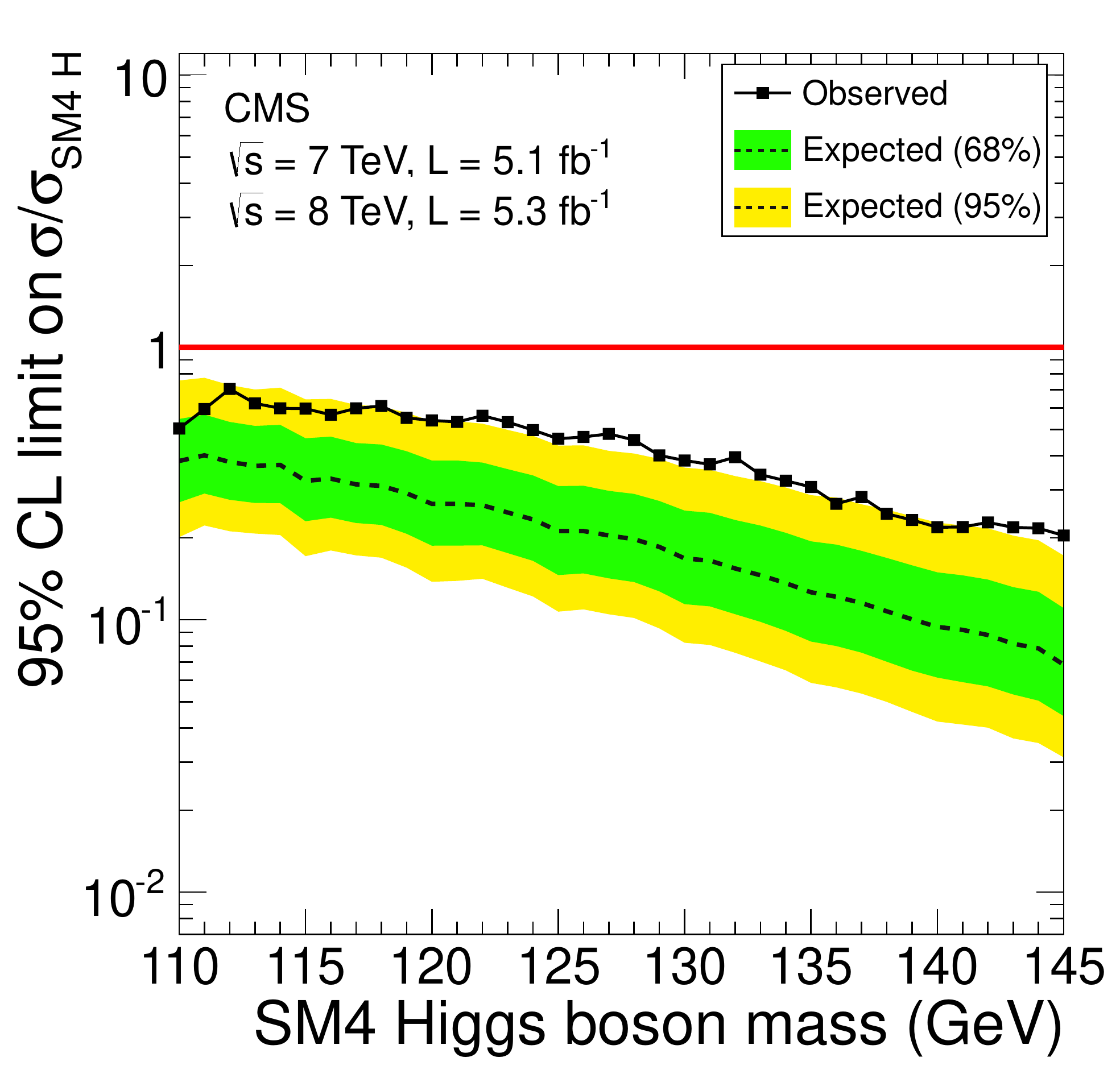}
\caption{
  The { observed and expected} 95\% CL upper limits on the signal strength modifier,
    $\mu = \sigma / \sigma_{\mathrm{SM4 \, H}}$, for the SM4 Higgs boson hypothesis as a function of the Higgs
  boson mass in the range 110--600\GeV (\cmsLeft) and 110--145\GeV (\cmsRight).
}
\label{fig:Mu95}
\end{figure}

The $\CLs$ value for the SM4 Higgs boson hypothesis as a function of its mass
is shown in Fig.~\ref{fig:CLs}.
$\CLs$ values of 0.05, 0.01, and 0.001 are indicated by horizontal thick red lines.
The mass regions where the observed $\CLs$ values are below these lines are excluded
with the corresponding ($1-\CLs$) confidence levels of 95\%, 99\%, and 99.9\%.
We exclude an SM4 Higgs boson
in the range \SMFourObsLNN--\SMFourObsHNN\GeV at 99\% CL,
and in the range \SMFourObsLNNN--\SMFourObsHNNN\GeV at 99.9\% CL.
Figure~\ref{fig:Mu95} shows the 95\% CL upper limits on the signal strength modifier,
$\mu = \sigma / \sigma_{\mathrm{SM4} \, \PH}$,
as a function of $\mH$.
The ordinate on this plot shows the Higgs boson cross section that is excluded
at 95\% CL, expressed as a multiple of the SM4 Higgs boson cross section.

\begin{figure} [phtb]
\centering
\includegraphics[width=\cmsFigWidth]{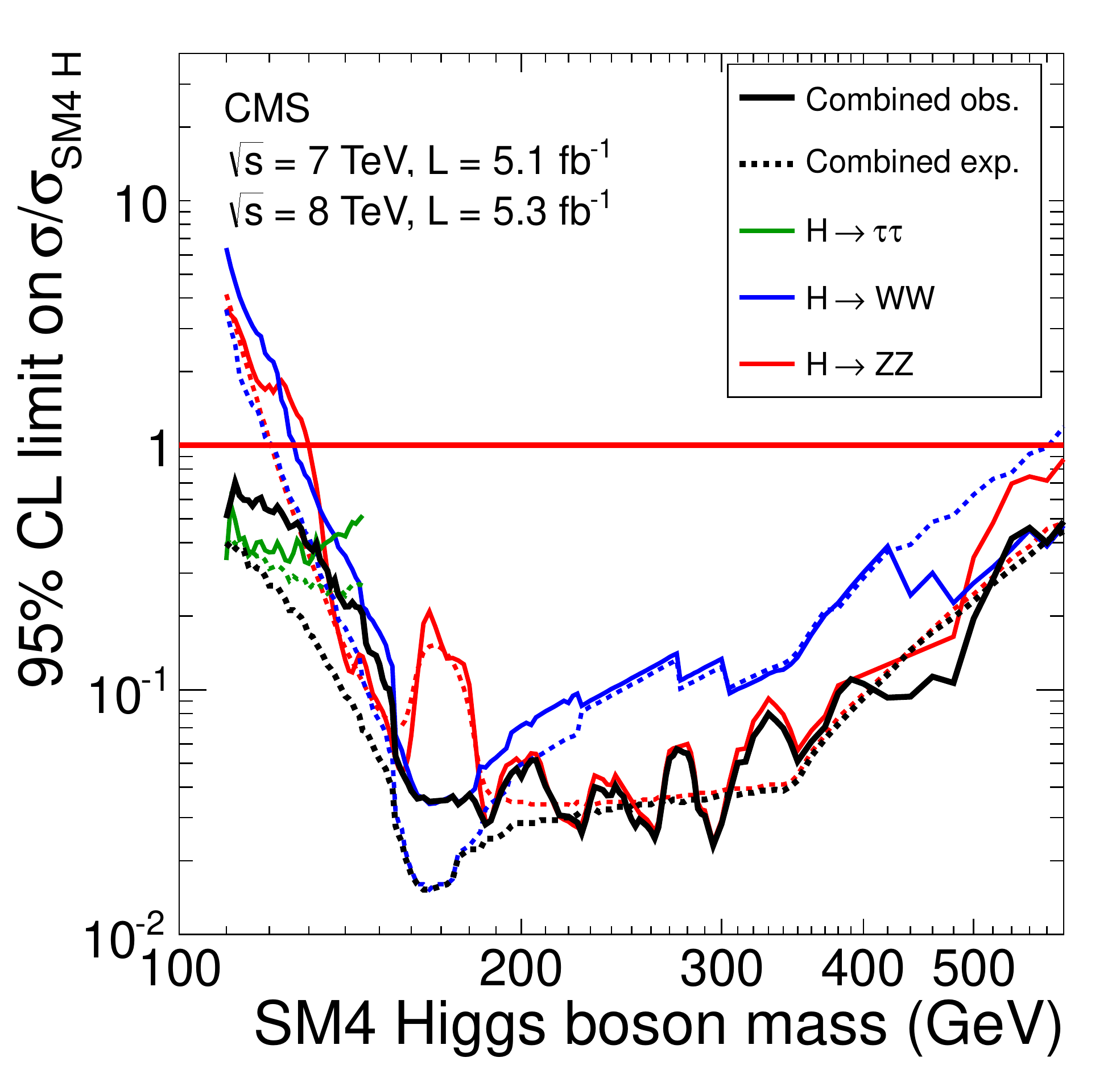}
\includegraphics[width=\cmsFigWidth]{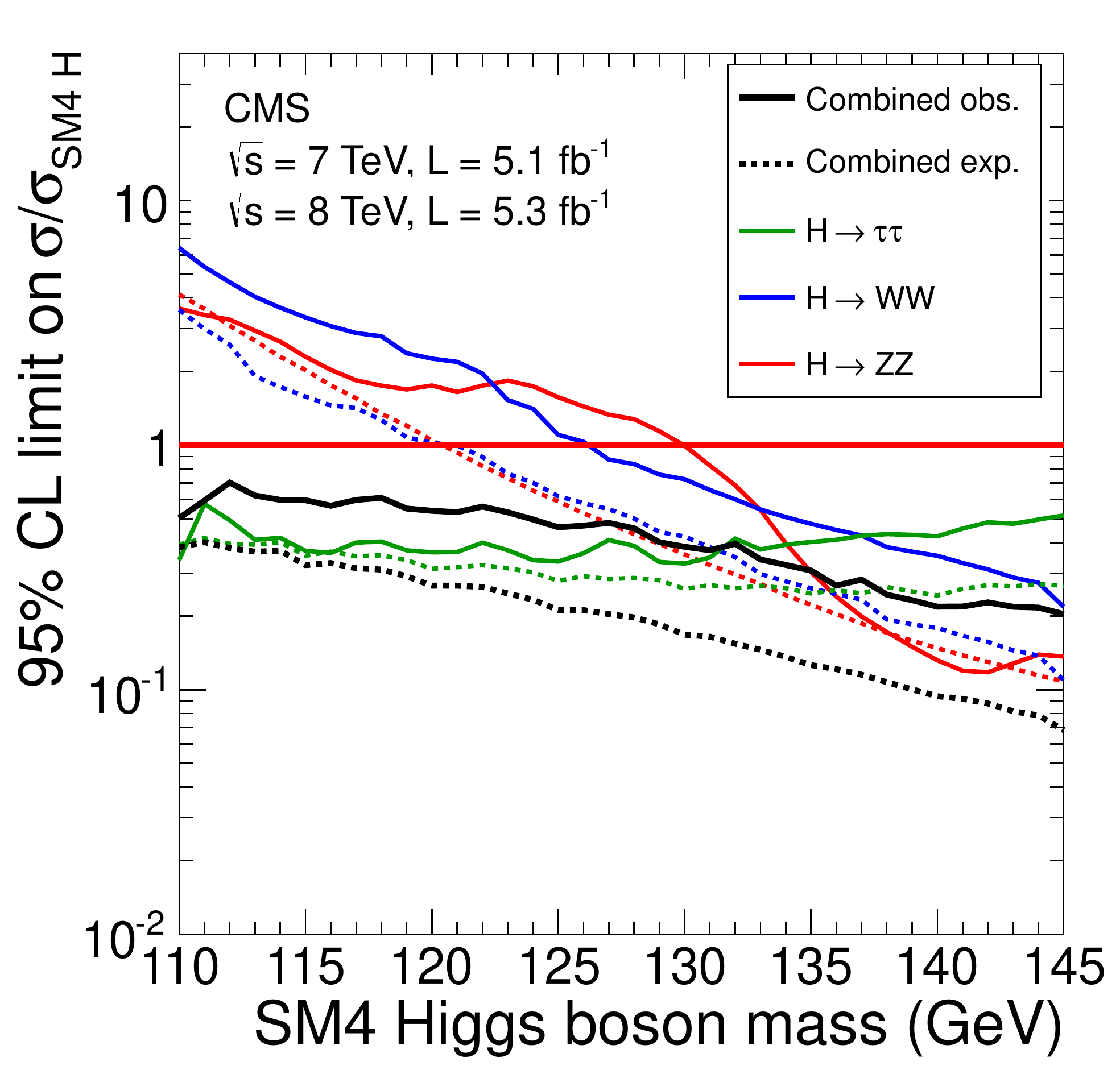}
\caption{
  The observed (solid lines) and expected (dashed lines) 95\% CL upper limits on the signal strength modifier,
    $\mu = \sigma / \sigma_{\mathrm{SM4 \, H}}$, as a function of the SM4 Higgs boson mass in the range 110--600\GeV
  (\cmsLeft) and 110--145\GeV (\cmsRight) for the { four explored} Higgs boson decay modes
  and their combination.
    }
\label{fig:Mu95_IndivChannels}
\end{figure}

Figure~\ref{fig:Mu95_IndivChannels} shows the observed and expected limits for
the three individual decay channels that have been considered, and their combination.
The $\PH \to \tau\tau$ search is
the most sensitive channel in the mass range below 135\GeV.
In the mass range 135--150\GeV, the best sensitivity is shared between
$\PH\to\cPZ\cPZ$ and $\PH\to\PW\PW$.
In the mass range 150--190\GeV, the $\PH\to\PW\PW$ channel has the best sensitivity.
For masses above 190\GeV, the
 sensitivity is driven mostly
by the $\PH\to\cPZ\cPZ$ decay channels.

\begin{figure} [htbp]
\centering
\includegraphics[width=\cmsFigWidth]{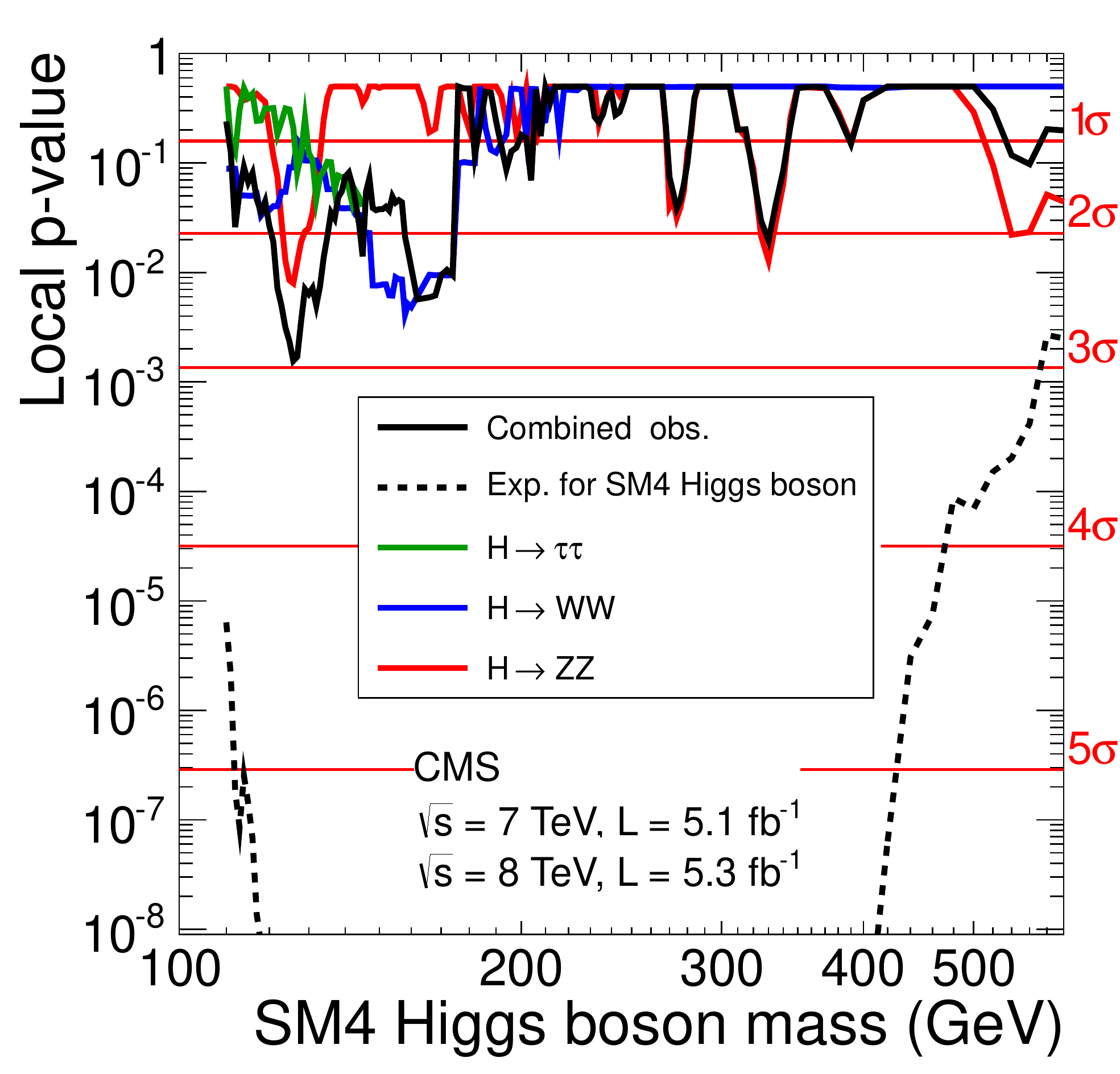}
\includegraphics[width=\cmsFigWidth]{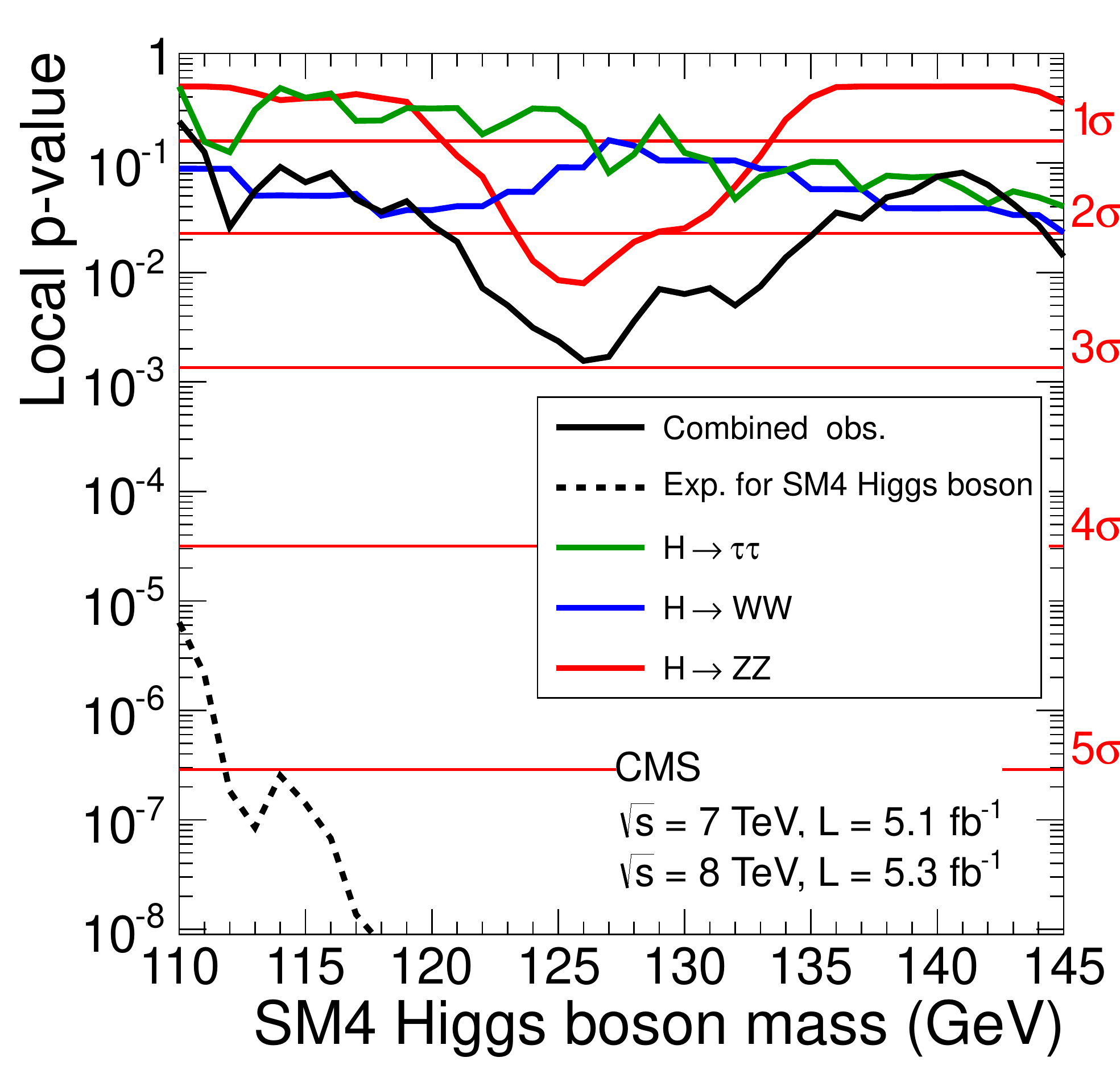}
\caption{
  The observed local $p$-value $p_0$
   as a function of the SM4 Higgs boson mass in the range 110--600\GeV (\cmsLeft) and 110--145\GeV (\cmsRight).
  The dashed line shows the expected local $p$-values
   should an SM4 Higgs boson with a mass $\mH$ exist.
    The expected $p$-value is obtained with nuisance parameters constrained by the data, giving it some dependence
     on the observed data, and hence the small modulations on top of the overall smooth trend as a function of $\mH$.
    }
\label{fig:pvalue}
\end{figure}

\begin{figure} [htbp]
\centering
\includegraphics[width=\cmsFigWidth]{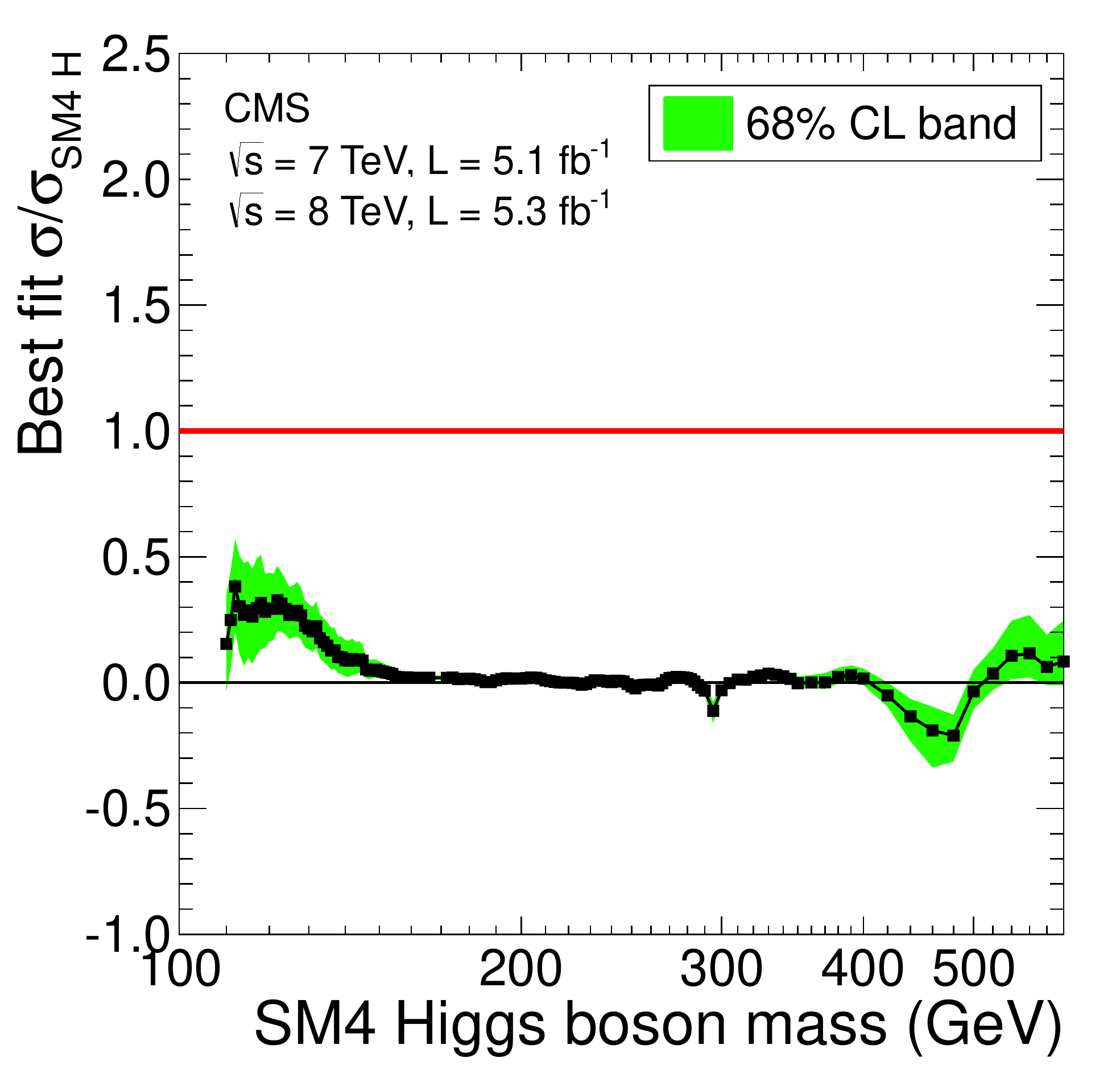}
\includegraphics[width=\cmsFigWidth]{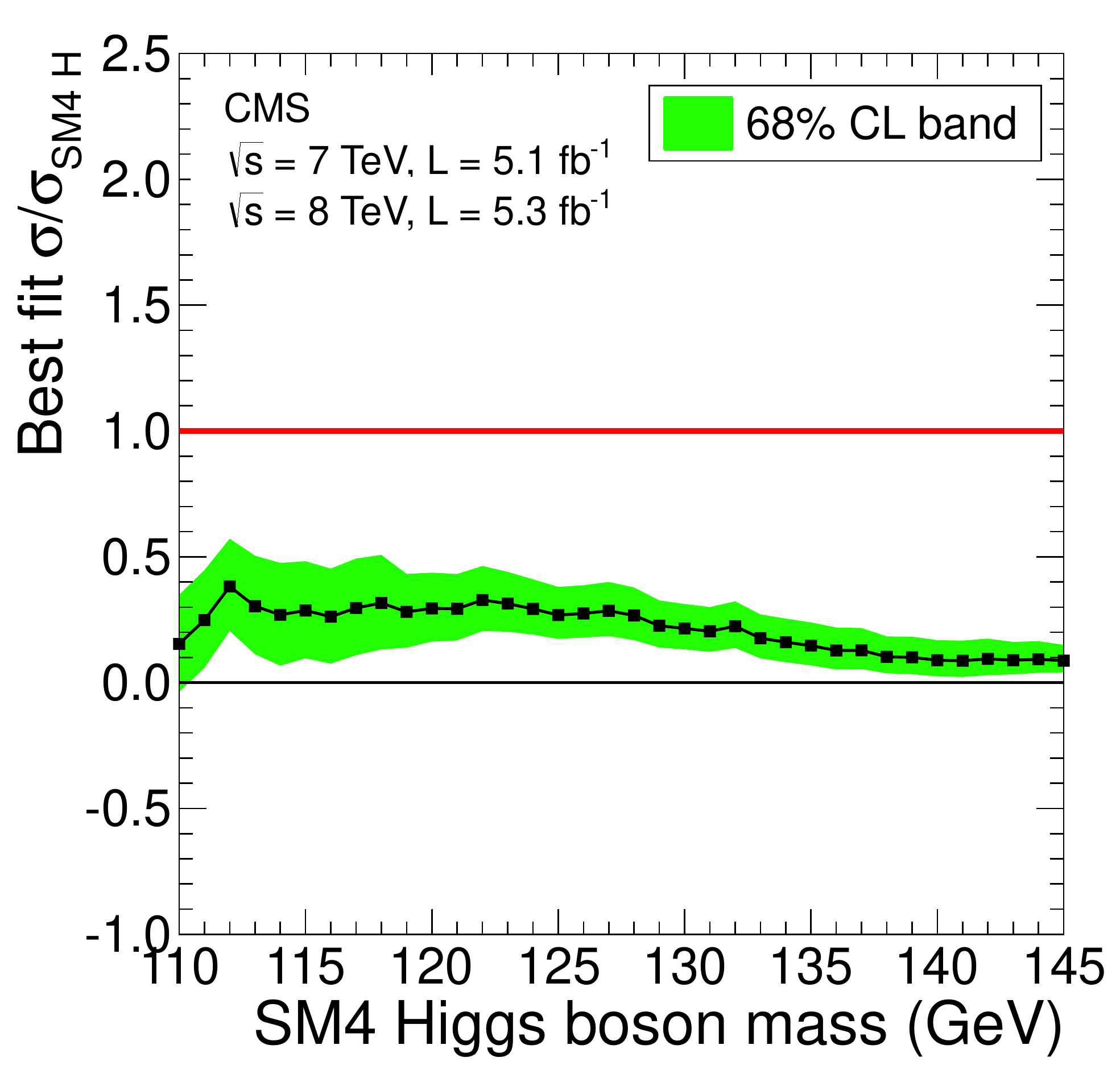}
\caption{
   The best-fit $\hat \mu = \sigma / \sigma_{\mathrm{SM4 \, H}}$
   as a function of the SM4 Higgs boson mass in the range 110--600\GeV (\cmsLeft) and 110--145\GeV (\cmsRight).
  The band corresponds to the $\pm 1$ standard deviation uncertainty on the $\hat \mu$ values.
    }
\label{fig:muhat}
\end{figure}

To quantify the consistency of the observed excesses
with the background-only hypothesis,
we show in Fig.~\ref{fig:pvalue}
a scan of the combined local $p$-value $p_0$, 
together with the results observed in the individual Higgs boson decay channels.
The minimum combined local $p$-value $p_{\text{local}}^{\text{min}}$ = \SMFourMinLocalP\
at $\mH \simeq \SMFourMaxZmass$\GeV
corresponds to a local significance $Z_{\text{local}}$ of $\SMFourMaxLocalZ \, \sigma$.
The global probability of observing at least as large an excess
somewhere in the entire search range 110--600\GeV is estimated
directly from the data using Eq.(\ref{eq:LEE1}).
The best-fit value $\hat \mu (\mH)$,
shown in Fig.~\ref{fig:muhat}, has
four upcrossings with $\hat \mu = 0$.
This can be better seen as upcrossings of the solid line above the dashed line in Fig.~\ref{fig:Mu95}.
Taking into account the number of observed upcrossings, the global $p$-value
of observing a local $\SMFourMaxLocalZ\,\sigma$ excess anywhere in the search region
for the background-only hypothesis is \SMFourGlobalPfull .

Figure~\ref{fig:muhat} also illustrates why the SM4 Higgs boson is excluded even though
a $\SMFourMaxLocalZ\,\sigma$ excess is observed at a mass near \SMFourMaxZmass\GeV.
The band shown in Fig.~\ref{fig:muhat} corresponds to
the $\pm 1$ standard deviation uncertainty (statistical+systematic) on the $\hat \mu$ value.
Given these uncertainties, the best-fit values of signal strength $\hat \mu (\mH)$
are significantly smaller than expected for the SM4 Higgs boson ($\hat \mu = 1$)
in the entire explored mass range.

Although the SM4 combination is not optimal for searching for the SM Higgs boson,
the presence of such a boson would
still produce an excess in the SM4 combination.
The expected significance for a SM Higgs boson with a mass near 125\GeV is $3.5 \, \sigma$,
which is very close to the observed value of $\SMFourMaxLocalZ \, \sigma$.
For reference, the expected significance at 125\GeV with the dedicated SM Higgs boson combination
is $5.8 \, \sigma$~\cite{CMSobsJul2012}.


\subsection{The FP results}
\label{sec:fp4results}

\begin{figure}[hbtp]
\begin{center}
      \includegraphics[width=\cmsFigWidth]{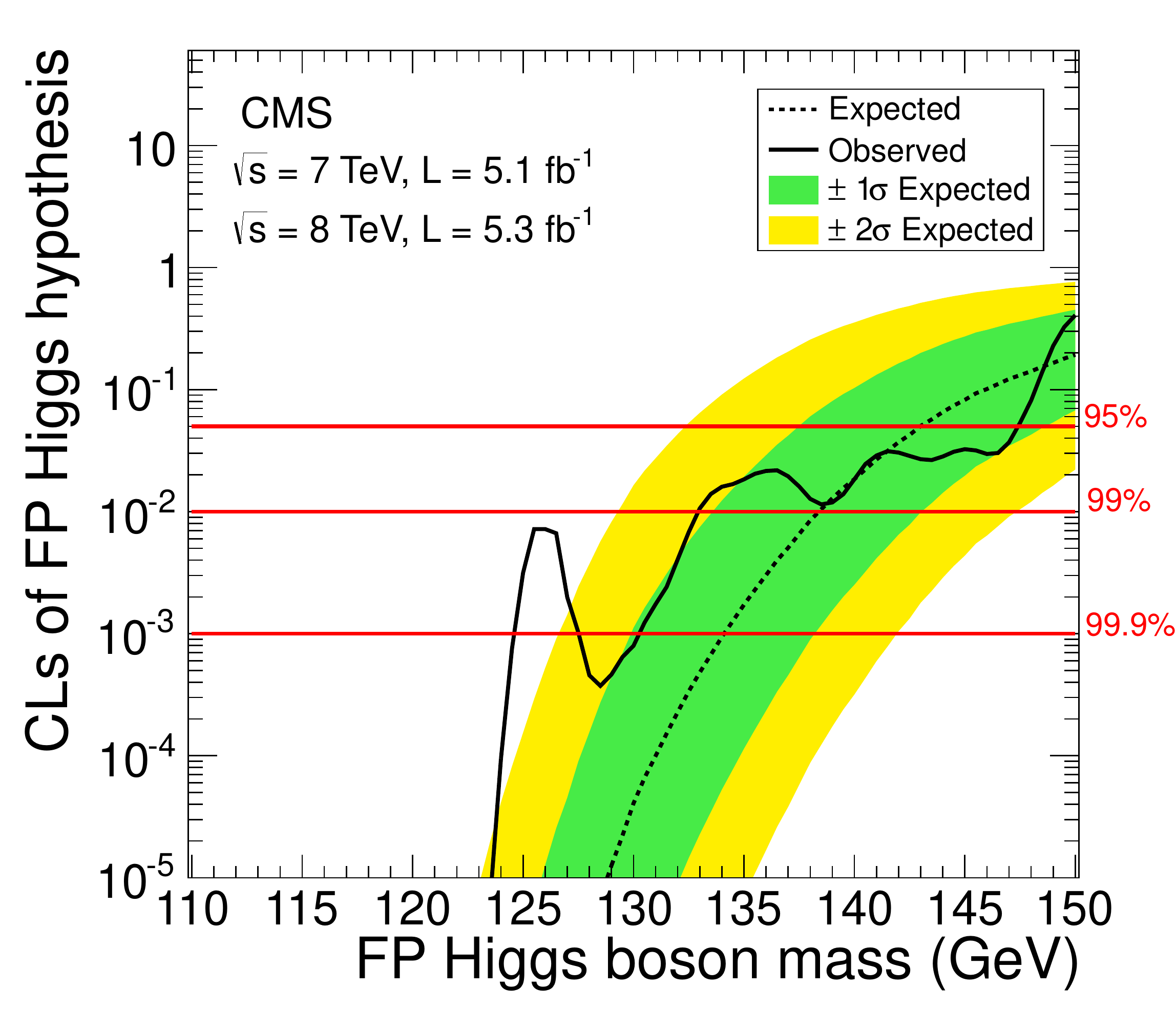}\hfill
      \includegraphics[width=\cmsFigWidth]{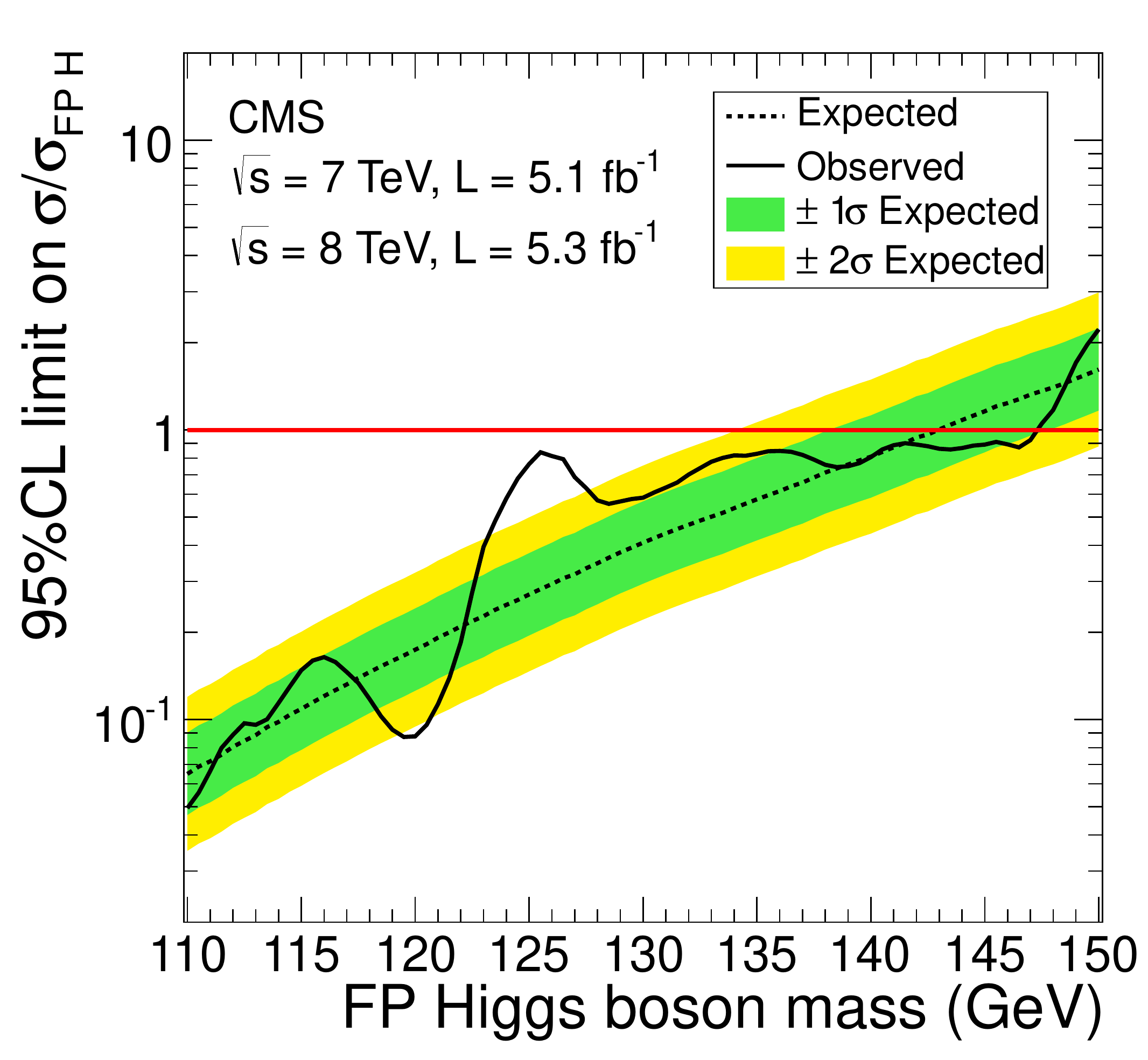}\hfill
        \caption{
	(\cmsLeft) The { observed and expected} $\CLs$ values for the FP Higgs boson hypothesis as a function
  of the Higgs boson mass in the range 110--150\GeV.
       (\cmsRight) The observed and expected 95\% CL upper limits on the signal strength modifier,
  $\mu = \sigma / \sigma_{\mathrm{FP} \, \PH}$, as a function of the FP Higgs boson mass in the range 110--150\GeV.
       \label{fig:CLs-limit}
             }
\end{center}
\end{figure}

The $\CLs$ value for the FP Higgs boson hypothesis as a function of its mass
is shown in Fig.~\ref{fig:CLs-limit} (\cmsLeft).
The $\CLs$ values of 0.05, 0.01, and 0.001 are indicated by thick red horizontal lines.
The mass regions where the observed $\CLs$ values are below these lines are excluded
with the corresponding ($1-\CLs$) confidence levels of 95\%, 99\%, and 99.9\%.
 The
fermiophobic Higgs boson is excluded at 95\% CL in the mass range 110--147\GeV and at 99\% CL in the range 110--133\GeV.

Figure~\ref{fig:CLs-limit} (\cmsRight)
shows the 95\% CL upper limits on the signal strength modifier,
$\mu = \sigma / \sigma_{\mathrm{FP}\,\PH}$, as a function of $\mH$.
The ordinate on this plot shows the Higgs boson cross section that is excluded at 95\% CL, expressed as a multiple of the FP Higgs boson cross section.

\begin{figure}[hbtp]
\begin{center}
      \includegraphics[width=\cmsFigWidth]{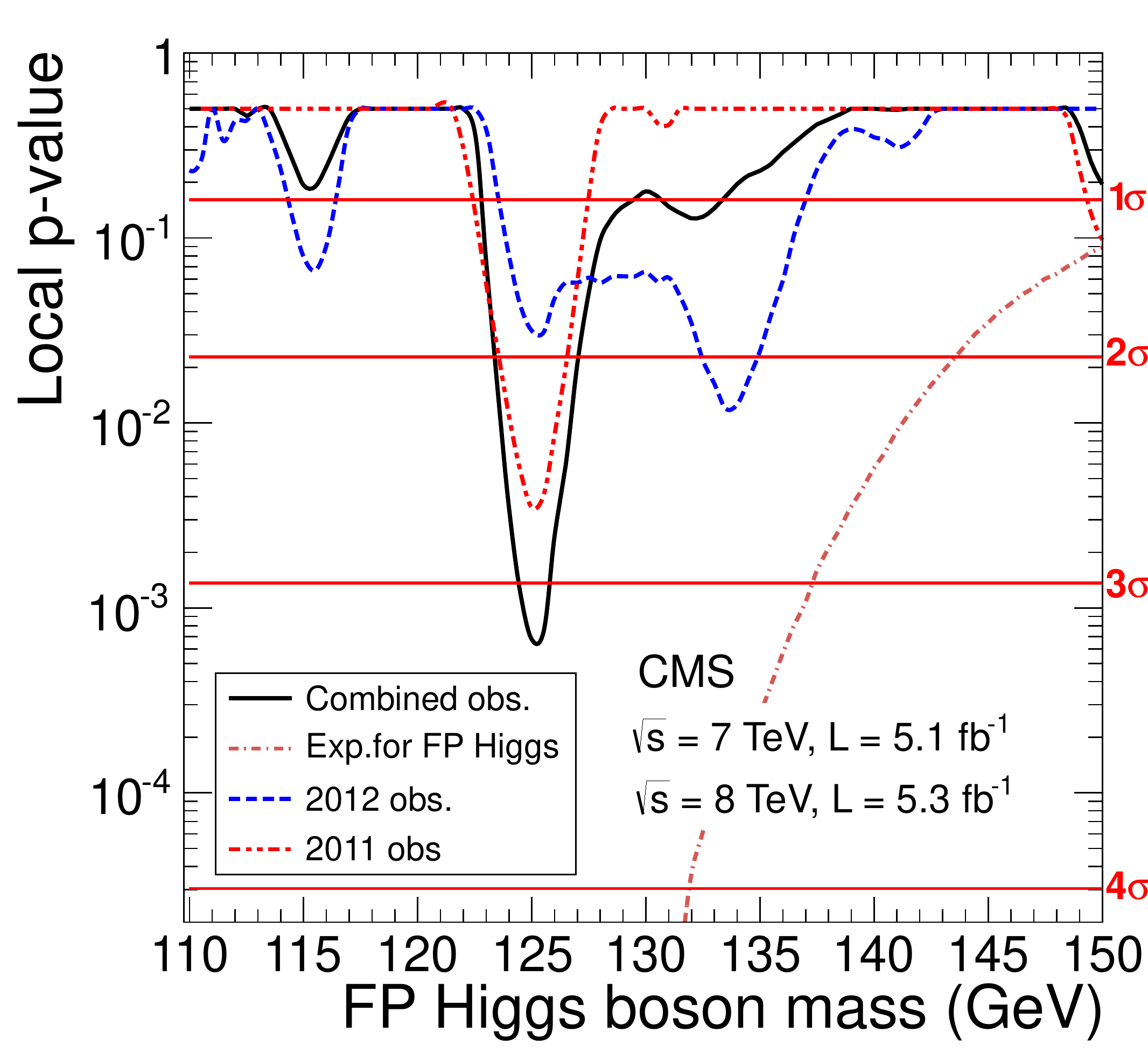}
      \includegraphics[width=\cmsFigWidth]{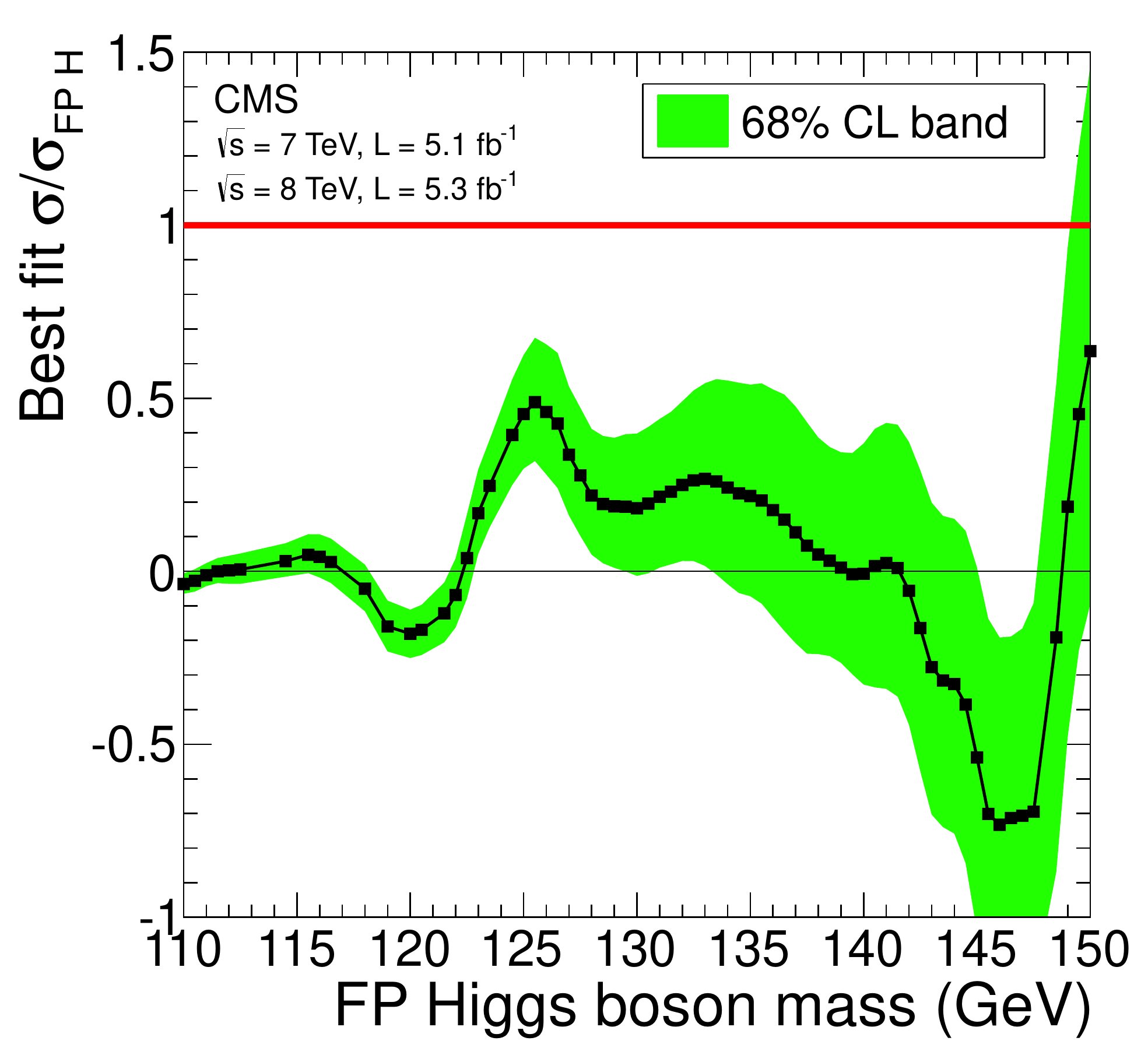} \\
        \caption{
        (\cmsLeft) The observed local $p$-value $p_0$
   as a function of the FP Higgs boson mass in the range 110--150\GeV. The dashed-dotted line shows the expected local $p$-values 
    should a fermiophobic Higgs boson with a mass
 $\mH$ exist. The contributions to the expected limit for each run period are shown.
	(\cmsRight) The best-fit $\hat \mu = \sigma / \sigma_{\mathrm{FP} \, \PH}$
   as a function of the FP Higgs boson mass in the range 110--150\GeV.
  The band corresponds to the ${\pm}1$ standard deviation uncertainty on the $\hat{\mu}$ values.
        \label{fig:Pvalue-MuHat}
             }
\end{center}
\end{figure}

Figure~\ref{fig:Pvalue-MuHat} (\cmsLeft) shows the local $p$-value as a function of the FP Higgs boson mass
for each run period and for their combination.
The largest upwards fluctuation of events over the expected background is observed at \FPMaxZmass\GeV,
and is computed to have a local significance of $3.2\,\sigma$.
This deviation from the expected limit
is too weak to be consistent with the fermiophobic Higgs boson signal,
as can be seen in Fig.~\ref{fig:Pvalue-MuHat} (\cmsRight), which shows
that the observed signal strength for a fermiophobic Higgs boson at \FPMaxZmass\GeV is $0.49\pm0.18$,
as obtained from the fit of signal plus background on data.
The excess of events at \FPMaxZmass\GeV is present
in the SM Higgs boson search reported
in Ref.~\cite{CMSobsJul2012}
and corresponds to the discovery of the new boson around 125\GeV.
This recently observed boson is not consistent
with a fermiophobic Higgs boson at 99\% confidence level.

As in the SM4 case, the FP analysis
is not optimal for searching for the SM Higgs boson, but still has some sensitivity.
The expected sensitivity to a SM Higgs boson with a mass of 125\GeV is $1.3\,\sigma$
we observe $3.2 \, \sigma$.
For reference, in the dedicated SM Higgs boson diphoton analysis, using the same dataset as
the FP combination here, the observed significance
of the excess near 125\GeV is $4.1 \, \sigma$,
with an expected  sensitivity of $2.8 \, \sigma$ ~\cite{CMSobsJul2012}.
In both the SM and FP diphoton analyses the observed significances
for the SM Higgs boson are greater than the expected,
but statistically compatible at the $\mathcal{O}(10\%)$ level.

\section{Summary}
Searches are reported for Higgs bosons in the context of either the standard model extended to include a fourth generation of fermions with masses of up to 600\GeV
or fermiophobic models. For the former, results from three decay modes ($\tau\tau$, $\PW\PW$, and $\cPZ\cPZ$) are combined, whilst for the latter the diphoton decay
is exploited.  The analysed proton-proton collision data correspond to integrated luminosities of up to {5.1}\fbinv at 7\TeV and
up to {5.3}\fbinv at 8\TeV. The observed results exclude the SM4 Higgs boson
in the mass range \SMFourObsLNN--\SMFourObsHNN\GeV at 99\% CL,
and in the mass range \SMFourObsLNNN--\SMFourObsHNNN\GeV at 99.9\% CL.
A fermiophobic Higgs boson is excluded in the mass range 110--147\GeV at 95\% CL, and in the range 110--133\GeV at 99\% CL.
The recently observed boson with a mass near 125\GeV is not consistent with either an SM4 or a fermiophobic Higgs boson.

\section*{Acknowledgements}
We congratulate our colleagues in the CERN accelerator departments for the excellent performance of the LHC and thank the technical and administrative staffs at CERN and at other CMS institutes for their contributions to the success of the CMS effort. In addition, we gratefully acknowledge the computing centres and personnel of the Worldwide LHC Computing Grid for delivering so effectively the computing infrastructure essential to our analyses. Finally, we acknowledge the enduring support for the construction and operation of the LHC and the CMS detector provided by the following funding agencies: BMWF and FWF (Austria); FNRS and FWO (Belgium); CNPq, CAPES, FAPERJ, and FAPESP (Brazil); MEYS (Bulgaria); CERN; CAS, MoST, and NSFC (China); COLCIENCIAS (Colombia); MSES (Croatia); RPF (Cyprus); MoER, SF0690030s09 and ERDF (Estonia); Academy of Finland, MEC, and HIP (Finland); CEA and CNRS/IN2P3 (France); BMBF, DFG, and HGF (Germany); GSRT (Greece); OTKA and NKTH (Hungary); DAE and DST (India); IPM (Iran); SFI (Ireland); INFN (Italy); NRF and WCU (Republic of Korea); LAS (Lithuania); CINVESTAV, CONACYT, SEP, and UASLP-FAI (Mexico); MSI (New Zealand); PAEC (Pakistan); MSHE and NSC (Poland); FCT (Portugal); JINR (Armenia, Belarus, Georgia, Ukraine, Uzbekistan); MON, RosAtom, RAS and RFBR (Russia); MSTD (Serbia); SEIDI and CPAN (Spain); Swiss Funding Agencies (Switzerland); NSC (Taipei); ThEPCenter, IPST and NSTDA (Thailand); TUBITAK and TAEK (Turkey); NASU (Ukraine); STFC (United Kingdom); DOE and NSF (USA). Individuals have received support from the Marie-Curie programme and the European Research Council (European Union); the Leventis Foundation; the A. P. Sloan Foundation; the Alexander von Humboldt Foundation; the Belgian Federal Science Policy Office; the Fonds pour la Formation \`a la Recherche dans l'Industrie et dans l'Agriculture (FRIA-Belgium); the Agentschap voor Innovatie door Wetenschap en Technologie (IWT-Belgium); the Ministry of Education, Youth and Sports (MEYS) of Czech Republic; the Council of Science and Industrial Research, India; the Compagnia di San Paolo (Torino); and the HOMING PLUS programme of Foundation for Polish Science, cofinanced from European Union, Regional Development Fund.

\bibliography{auto_generated}   

\cleardoublepage \appendix\section{The CMS Collaboration \label{app:collab}}\begin{sloppypar}\hyphenpenalty=5000\widowpenalty=500\clubpenalty=5000\textbf{Yerevan Physics Institute,  Yerevan,  Armenia}\\*[0pt]
S.~Chatrchyan, V.~Khachatryan, A.M.~Sirunyan, A.~Tumasyan
\vskip\cmsinstskip
\textbf{Institut f\"{u}r Hochenergiephysik der OeAW,  Wien,  Austria}\\*[0pt]
W.~Adam, E.~Aguilo, T.~Bergauer, M.~Dragicevic, J.~Er\"{o}, C.~Fabjan\cmsAuthorMark{1}, M.~Friedl, R.~Fr\"{u}hwirth\cmsAuthorMark{1}, V.M.~Ghete, J.~Hammer, N.~H\"{o}rmann, J.~Hrubec, M.~Jeitler\cmsAuthorMark{1}, W.~Kiesenhofer, V.~Kn\"{u}nz, M.~Krammer\cmsAuthorMark{1}, D.~Liko, I.~Mikulec, M.~Pernicka$^{\textrm{\dag}}$, B.~Rahbaran, C.~Rohringer, H.~Rohringer, R.~Sch\"{o}fbeck, J.~Strauss, A.~Taurok, W.~Waltenberger, G.~Walzel, E.~Widl, C.-E.~Wulz\cmsAuthorMark{1}
% ...
\vskip\cmsinstskip
\dag:~Deceased\\
1:~~Also at Vienna University of Technology, Vienna, Austria\\
% ...

\end{sloppypar}
\end{document}